\documentclass[preprint2]{aastex}
\usepackage{times,url,graphicx,amssymb}

\newcommand{\Ha}{\mbox{H$\alpha$}}
\newcommand{\Hb}{\mbox{H$\beta$}}

\shorttitle{Structure and Feedback in 30 Doradus}
\shortauthors{Pellegrini, Baldwin \& Ferland}

\begin{document}

\title{Structure and Feedback in 30 Doradus I: Observations }

\author{E.W. Pellegrini\altaffilmark{1,2}}
\affil{Department of Astronomy, University of Michigan, 500 Church Street, Ann Arbor, MI 48109}
\author{J.A. Baldwin}
\affil{Physics and Astronomy Department, Michigan State University, 3270 Biomedical Physical Sciences Building, East Lansing, MI 48824, USA}
\author{G.J. Ferland}
\affil{Department of Physics and Astronomy, University of Kentucky, 177 Chemistry /Physics Building, Lexington, KY 40506, USA}
\email{pelleger@umich.edu}

\altaffiltext{1}{Visiting astronomer, Cerro Tololo Inter-American Observatory, National Optical Astronomy Observatory, which are operated by the Association of Universities for Research in Astronomy, under contract with the National Science Foundation.} 
\altaffiltext{2}{Physics and Astronomy Department, Michigan State University, 3270 Biomedical Physical Sciences Building, East Lansing, MI 48824, USA} 

\begin{abstract}
We have completed a a new optical imaging and spectrophotometric survey of a 140 x 80 pc$^2$ region of 30 Doradus centered on R136, covering key optical diagnostic emission lines including \Ha, \Hb, H$\gamma$, [O~III] $\lambda\lambda$4363, 4959, 5007, [N~II] $\lambda\lambda$6548, 6584, [S~II] $\lambda\lambda$6717, 6731 [S~III] $\lambda $6312 and in some locations [S~III] $\lambda$9069. We present maps of fluxes and intensity ratios for these lines, and catalogs of isolated ionizing stars, elephant-trunk pillars, and edge-on ionization fronts. The final science-quality spectroscopic data products are available to the public. Our analysis of the new data finds that, while stellar winds and supernovae undoubtedly produce shocks and are responsible for shaping the nebula, there are no global spectral signatures to indicate that shocks are currently an important source of ionization. We conclude that the considerable region covered by our survey is well described by photoionization from the central cluster where the ionizing continuum is dominated by the most massive O stars. We show that if 30 Dor were viewed at a cosmological distance, its integrated light would be dominated by its extensive regions of lower surface-brightness rather than by the bright, eye-catching arcs.\\
\end{abstract}

\section{Introduction}
A significant share of our knowledge about star-formation rates, chemical abundances and abundance gradients in galaxies comes from studying emission lines from distant Giant Extragalactic H\ II Regions (GEHRs). But the observations of GEHRs are generally interpreted in the absence of a quantitative understanding of the structure responsible for the emission, or of its relationship to the processes of star formation.  Massive stars heat, fragment and compress the gas clouds from which they have recently formed. This ``stellar feedback'' is believed to drive the collapse of gravitationally bound clouds and trigger ongoing star formation (Zavagno et al. 2010; Oey et al. 2005; Elmegreen \& Lada 1977) and also shapes the structure of the gas which in turn affects the emission line spectrum. For these reasons, it is important to understand how such processes work.

Most GEHRs are distant, with poor spatial resolution. What is needed to better understand the mechanisms of stellar feedback is a nearby ``Rosetta stone'' in which we can resolve spatial details on the scale over which the feedback effects operate, for example over the pc-scale thickness of the ionized layer in an H II region. 30 Dor in the Large Magellanic Cloud, a nearby star-forming region that can be classified as a GEHR, is that key example. It is the largest star-forming region in the Local Group. The central cluster emits almost 500 times more ionizing photons than the Orion Nebula and 2--4 times more ionizing photons than other large Local Group star-forming regions such as NGC 3603 in our Galaxy or NGC 604 in M33. Because 30 Dor is only 48.5 kpc away (Macri et al. 2006), it can be studied with high spatial resolution (1'' = 0.25 pc). 

The violent history of 30 Dor has led to the formation of vast cavities and shell-like structures around its central star cluster NGC~2070 (Meaburn 1984; Selman et al. 1999). The walls of the cavities have long been seen in optical ionization-front tracers like [S~II]. The SAGE ~\citep{Meixner2006} survey of the LMC included 30 Dor. SAGE traces the ionization-fronts in 8$\micron$\ PAH emission with a resolution of 5 arcsec. Filling these cavities is a $10^6$ K gas visible in X-ray emission ~\citep{Townsley2006}. Many of these structures have \Ha~ velocity profiles which show them to be expanding shells ~\citep{Meaburn1984}. These shells have typical kinetic energies equal to $10^{51}$ erg with a total kinetic energy in the vicinity of NGC~2070 equal to $10^{52}$ erg ~\citep{ChuKennicutt1994}. These types of kinematic arguments suggest SNe are a major component of the total energy budget. However, within reasonable estimates about the timescale for injecting energy from stellar winds into the ISM it is equally likely that stellar winds are the agent shaping the nebula.
 
The remnants of the molecular gas cloud from which the stars have formed has been observed in optically thick CO emission ~\citep{Poglitsch1995}. This molecular cloud runs NE-SW behind NGC~2070. The brightest optical arcs, seen in Figure \ref{fig:HaMosaic}, appear to trace the surface of this molecular cloud, and are bright because of their high density. 30 Dor has been named the ``Tarantula Nebula'' because images showing emission lines are dominated by the interwoven pattern of these bright arcs. However, as we will show here, the majority of the line emission originates in the extended regions of low surface brightness. 

This current paper describes our new optical-passband imaging and spectroscopic survey of 30 Dor. Our data set, which is publicly available, is designed to include key emission lines which quantitatively describe feedback processes. One of our goals is to survey the full region on the sky that would be lumped together in observations of a distant GEHR, for direct comparison to such objects. Luminous H\ II region complexes in distant galaxies typically are about 100 pc in diameter (Oey et al. 2003; Hunt \& Hirashita 2009).  The 140 $\times$ 80 pc$^2$ area covered by our spectra is of similar size, and encompasses 50 \% of the \Ha~ flux from the 10' diameter region with \Ha~ surface brightness appreciably above the general background level in this part of the LMC as measured from the Magellanic Cloud Emission Line Survey (MCELS; ~\citet{Smith1998}). The larger 11' $\times$ 12' area of our sub-arcsec resolution imaging survey encompasses 80 \% of the total \Ha~ flux.  

In 30 Dor, the numerous expanding shells are caused by mechanical energy input (Meaburn 1984; Chu \& Kennicutt 1994), and the ionizing energy source for a radius of at least 100 pc is generally accepted to be photoionization by the central cluster NGC 2070 (Elliott et al. 1977; Tsamis et al. 2003). However, recent work based on Spitzer observations (Indebetouw et al. 2009; hereafter I09) has questioned this. In this paper we use our observations to investigate the source of the photons that ionize the gas beyond the bright arcs. 

The ionization and shaping need not be caused by the same mechanism. The application of our new data to studying feedback processes -- the question of what has shaped the nebula -- will be presented in a companion paper (Pellegrini et al. 2010, hereafter Paper II). There we will also discuss the consequences of the resulting structure on the escape of ionizing radiation. Finally, the second paper will evaluate the results from ``strong line techniques'' (Pagel et al. 1979; Kewley \& Doptita 2002; Denicol{\'o} et al. 2002) applied to our composite spectrum of 30 Dor, by comparing them to the chemical abundances and other nebular properties obtained by a point-by-point analysis of our spatially resolved dataset.

\section{Observations}
\subsection{Existing optical passband data sets}
Most of the strong optical emission lines from star-forming regions trace ionized gas at electron temperatures $T_e \sim 10^4$ K. A few key lines from [O~I], [O~II], [N~II] and [S~II] form in the interface region at the edge of ionization-bounded H\ II regions. The elements S, O, N, Ar have transitions visible from the ground that are sensitive to both $T_e$ and the electron density $n_e$, and also to gas-phase abundances. Even though 30 Dor is a key example of a GEHR, the available measurements of it in the emission lines of these elements, using either direct imaging or especially spectroscopy, are surprisingly limited.

The best existing publicly available narrow-band optical image data sets are the MCELS, and archival HST images. The MCELS survey covers the central 8$\times$8 deg$^2$ of the LMC including 30 Dor, in \Ha~ + [N~II], [O~III] and [S~II] (using many of the same emission line and continuum filters used in this present study). However, the data were taken with the 0.9m Schmidt telescope at CTIO and have a spatial scale of 2.3 arcsec pixel$^{-1}$ with a resolution closer to 5 arcsec FWHM. This means that structures smaller than 1.2 pc are unresolved, blurring fine details.

The HST archival data cover the central 4.5 by 3 arcmin of 30 Dor with at least 0.1 arcsec resolution (Scowen 1998; Walborn et al. 2002) in \Ha, [O~III] and [S~II], revealing vastly more detail. However the HST data are limited to the brighter central region around R136, the very compact group of stars at the very center of NGC~2070. Since the brighter nebula accounts for only 25 per cent of the total nebular emission in \Ha~ the limited spatial coverage of HST misses the bulk of the emission that would be detected if 30 Dor were viewed from a much greater distance.

Turning to the available spectroscopy, ~\citet{KrabbeCopetti2002} obtained a set of long-slit observations which covered \Hb, [O~III] $\lambda$4363 and [O~III] $\lambda$5007 at 135 points along three slit positions. The sensitivity of these spectra is comparable to our Blanco survey described below.  
~\citet{MCP1985} had previously obtained spectra covering a similar wavelength range at four other slit positions. They presented line strengths measured for 22 extracted regions covering typically 30 to 60 arcsec each along the slit.

~\citet{ChuKennicutt1994} used the CTIO echelle spectrograph in a single-order mode to obtain a sparse grid of long-slit spectra that covered the \Ha~ and [N~II] $\lambda\lambda$6548,6584 emission lines with high (20 km s$^{-1}$) velocity resolution. They found that about half of the kinetic energy in 30 Dor is contained in shells in the central regions which are expanding with characteristic velocities $v \sim 20-200$ km s$^{-1}$, and that the kinetic energy contained in this expansion greatly exceeds the gravitational binding energy.

There has also been a limited amount of deep echelle spectroscopy covering a much wider wavelength range (Tsamis et al. 2003; Peimbert 2003). These spectra measured hundreds of emission lines that can be used for detailed chemical abundance analysis, and also cleanly resolve the density-sensitive [O~II] $\lambda\lambda$3726,3729 doublet and many other lines which are blended at lower spectral resolution. However, they require relatively long exposures and very short slit lengths, so have only very small areal coverage on 30 Dor. 
These existing measurements will be discussed below as needed.

\subsection{New narrow-band images}

To the existing data sets we added a new set of narrow-band images taken with the SOAR Optical Imager (SOI) on the 4.1m SOAR Telescope\footnote{The Southern Astrophysical Research (SOAR) Telescope is a joint project of Michigan State University, Minist\'{e}rio da Ci\^{e}ncia e Tecnologia-Brazil, the University of North Carolina at Chapel Hill, and the National Optical Astronomy Observatory. Further information about SOAR and its instruments may be found at www.soartelescope.org.}.  We used the \Ha~ 6563/75, [S~II] 6738/50, [O~III] 5019/15 emission-line filters and the 6850/100 and 5130/155 continuum filters from the CTIO 3$\times$3 and 4$\times4$ in$^2$ filter sets, where the filter names refer to the approximate central wavelengths and FWHM bandpasses.  A summary of the SOI observations is given in Table \ref{tab:SOIsummary}.

In each passband we took grids of 5$\times$5 arcmin$^2$ SOI images, overlapping them to give a 12$\times$13 arcmin$^2$ field of view with a scale of 0.15 arcsec pixel$^{-1}$. The individual images were then combined to create final mosaic images in each passband. Due to the significant spatial overlap of the individual images, the total integration time in a given filter varies across the mosaic. Table \ref{tab:SOIsummary} includes, for each filter, the resulting minimum and maximum integration times at any point in the mosaic. The data are seeing-limited, with full width at half maximum intensity FWHM = 0.5 - 0.9 arcsec, much better than the 5 arcsec FWHM in the MCELS survey, and are three times more sensitive to diffuse nebular emission than the existing HST data set.

\begin{figure*}
\center{\includegraphics[width=\textwidth]{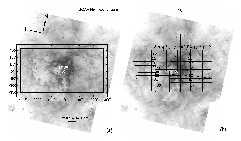}}
\caption{New narrow band \Ha~ image of 30 Dor taken with the SOAR telescope. North is rotated 13 deg clockwise from vertical. The center of R136 is marked as a white ``$\times$'' symbol. (a) The outline of the region covered by our maps made from the Blanco spectra. (b) The individual slit positions of our Blanco spectroscopic data set.}
\label{fig:HaMosaic}
\end{figure*}

As an example of the results, Figure \ref{fig:HaMosaic}a shows the mosaic image made with the \Ha + [N~II] filter, prior to continuum subtraction. R136 is marked with a white $\times$. The 1 arcmin scale bar is equivalent to 14.1 pc for an LMC distance of 48.5 kpc ~\citep{Macri2006}. Figure \ref{fig:HaMosaic}b shows, on the same image, the slit positions used in the SOAR Telescope spectroscopic observations described below.
  
The two continuum filters were used to measure the emission from stars, together with the nebular continuum and starlight scattered off dust. The continuum image count rates were scaled and subtracted from the emission-line image to give a nebular count rate $R_{line}$ in each filter, according to
\begin{equation}
\label{eq:rate}
R_{line} = R_{narrow} - R_{cont} \frac{W_{narrow}}{W_{cont}}
\end{equation}
 where $R_{narrow}$ and $R_{cont}$ are equal to the atmospheric-extinction corrected count rate in the narrow band and continuum filters, respectively, and the effective filter width $W_{i}$ in the $i^{th}$ filter is
\begin{equation}
W_i = \int{T_i(\lambda)d\lambda}
\end{equation}
where the transmission curve $T_i(\lambda)$ was measured by the SOAR staff.

Changes in stellar scattered light, seeing or sky brightness between the time the emission lines were measured and the continuum was measured were found to be a limiting factor in the quality of the continuum-subtracted images, so the observations with the continuum and emission filters were made sequentially one after the other before offsetting the telescope to each new point in the mosaic grid. This ensures that the relative flux between emission line and continuum images remains constant for each pointing.
\subsection{Spectrophotometry}
\subsubsection{The Blanco Telescope Spectral Grid}
In Feb 2008 we obtained a grid of long-slit spectra of 30 Dor using the RC spectrograph on the 4m Blanco telescope at Cerro Tololo Interamerican Observatory (CTIO).  The spectrograph slit was 5 arcmin in length, and was positioned at a total of 37 different locations spanning the nebula (Figure  \ref{fig:HaMosaic}b). Two sets of spectra were taken at orthogonal angles. The slit position angles PA = 13 deg and PA = 103 deg were chosen to maximize the number of key ionization fronts that could be covered with the slit either crossing the ionization fronts at an approximately perpendicular angle or running directly along them. The Blanco telescope uses an equatorial mount, which results in a constant position angle of the slit once the instrument has been rotated to the correct PA. We minimized the uncertainty in the PA by rotating the slit only once each night, after all observations at the initial PA were completed. As a result we have a high degree of confidence in our stated position angles, and therefore of the mapping onto the sky of individual points along the slit.

Table \ref{tab:SPECpointings} lists for each slit position the identifying position number, the RA and Dec of the slit center in J2000 coordinates, the PA, and the total exposure time. For simplicity, slit positions with PA = 13 deg are numbered beginning at 1 at the easternmost position and increasing to 17 toward the west. These include the two most extreme north and south slit positions (position numbers 16 and 17, respectively). Numbering of slit positions with PA =103 deg then begins at 20 and increases towards the south, with even numbered slit positions centered to the east of R136 and odd numbered ones to the west. The only exception is position 31 which has a PA equal to 98 degrees. Positions 40-44 were taken with the SOAR Telescope and are described in the following sub-section.

The Blanco Telescope observations were taken under photometric conditions with a slit width of 3.5 arcsec sampled at 0.5 arcsec/pixel, using the 600 l/mm grating. This setup delivers a spectral resolution of 7.0\AA\ (FWHM) with a spectral sampling of 1.06 \AA/pixel as measured from the [O\ I] $\lambda$5577.33 \AA\ night sky line. Accurate offsetting was achieved by moving the autoguider probe by known amounts, and fiducial stars of known coordinates were placed in the slit at most of the slit positions in order to internally calibrate the position on the sky. An autoguider maintained the position to better than 1 arcsec.

The standard IRAF 'LONGSLIT' package was used to generate bias and flat-field corrected, flux calibrated long-slit spectra, rectified to remove optical distortions and to convert to a linear wavelength scale. The wavelength solution had an RMS uncertainty of 0.06 \AA.  After performing the distortion correction using the wavelength calibration, any additional wavelength shifts due to flexure within the spectrograph were corrected by shifting each spectrum so that the [O I] $\lambda$5577.33 \AA\ night sky line fell at the correct wavelength. The spectrophotometric standard stars LTT 2415, Hiltner 600 and CD32 from ~\citet{Hamuy1994} were observed on each night, to obtain an absolute flux calibration. A 7 arcsec slit was used for the flux calibrations to ensure that no light was lost in the wings of the stellar profile.

\subsubsection{Additional SOAR Telescope Spectroscopy}
Six additional locations in 30 Dor were observed with the Goodman High Throughput Spectrograph ~\citep{Clemens2004} on the SOAR telescope, on 5 Feb 2009. The instrument was used with a single slit 3.9 arcmin long and 0.46 arcsec wide, with its 300 line mm$^{-1}$ grating. The wavelength range was 3950 \AA\ to 9335 \AA\ with 1.32 \AA\ pixel$^{-1}$ sampling and a spectral resolution of 4.9 \AA\ FWHM at \Ha. This setup easily resolved the [S~II] doublet near 6720 \AA\ and the [O~II] lines at $\lambda\lambda$7320, 7330 \AA.

Calibration of the data was performed with the spectroscopic packages in IRAF in a fashion similar to the method described above for the Blanco data, with a few notable exceptions. The spectrophotometric standard stars LTT 2415 and LTT 4816 were observed, without a slit, to obtain an absolute flux calibration. There is second-order contamination in the spectra beyond 7700 \AA, with the sensitivity in second order below 3850 \AA\ being about 36 per cent of that in first order at the same wavelength. This is not a major problem for measurements of the nebular emission lines in the red (observed in first order), because at the positions of these lines the contaminating second-order light is mostly continuum emission. But the second-order contamination does strongly affect the flux calibration.

Second-order contamination in the red spectrum of the standard stars was accounted for by observing the standard stars through two different second-order blocking filters, GG385 and GG495, which cut off light blueward of 3850 and 4950 \AA, respectively. The GG385 filter was used for the nebular observations and for the flux calibration for wavelengths lower than 7000 \AA, while the GG495 filter was used only for the flux calibration beyond 7000 \AA. To stitch together a flux calibration combined over the full wavelength range, a sensitivity response curve was generated separately for each filter, using the IRAF 2D spectroscopic reduction package.  A final response curve was made using the GG 385 curve below 7000 \AA\ and the GG495 curve at longer wavelengths.

Fringing with the current CCD is severe, reaching 32 per cent at 9000 \AA. To calibrate this, an internal quartz lamp was observed immediately after the series of nebular exposures at each slit position, before moving the telescope.  In spite of this precaution, flexure within the instrument still caused significant, obvious offsets between the fringes in the nebular data and in the quartz calibration frame. We used the fringe pattern in the spectra of fiducial stars that fell in the slit as a guide for shifting and scaling in amplitude the quartz fringe frames before dividing out the fringe pattern. This procedure decreased the amplitude of the residual fringing to 10 percent.

\begin{figure*}
\center{\includegraphics[width=\textwidth]{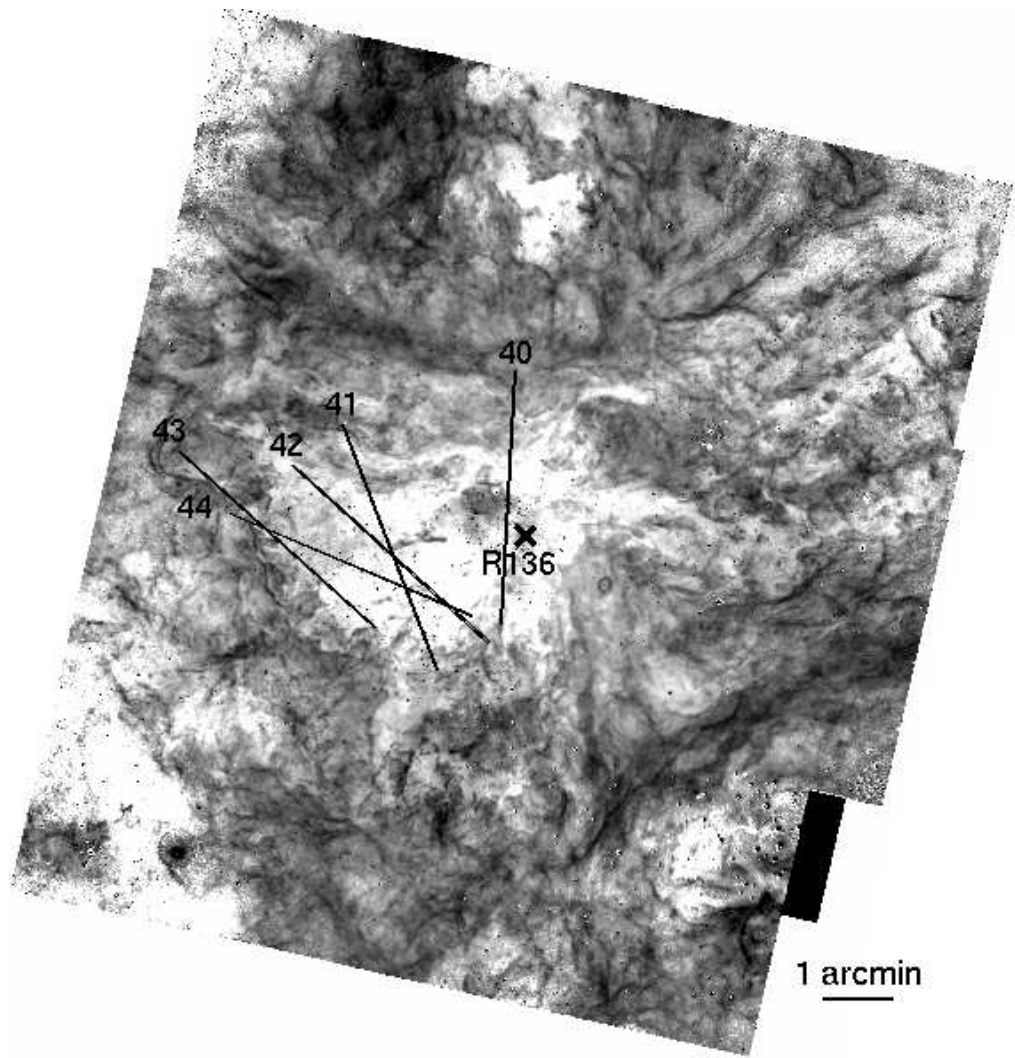}}
\caption{Ratio of [S\ II]/\Ha\  from our SOAR narrow-band images. Darker shades indicate a higher ratio.  The orientation is the same as Figure \ref{fig:HaMosaic}. For reference, the SOAR spectroscopic slit positions are plotted on top of the image.}
\label{fig:S2Ha_ratio}
\end{figure*}

The SOAR spectra sample an area of very low [S~II]/\Ha~ intensity ratio, lying to the east of R136. Figure \ref{fig:S2Ha_ratio} shows the SOAR slit positions superimposed on a map of the [S~II]/\Ha~ intensity ratio made from our SOAR images. The orientation and spatial coverage of Figure \ref{fig:S2Ha_ratio} is the same as that of Figure \ref{fig:HaMosaic}. At the time of the observations the SOAR Telescope lacked an atmospheric dispersion corrector, which is important with our broad wavelength coverage and narrow slit. As a result our observations were restricted to nearly the parallactic angle, resulting in the pattern shown in Figure \ref{fig:S2Ha_ratio}. The positions of the SOAR spectra are numbered beginning at 40, to minimize confusion between the two spectroscopic data sets.

\begin{figure*}
\center{\includegraphics[width=\textwidth]{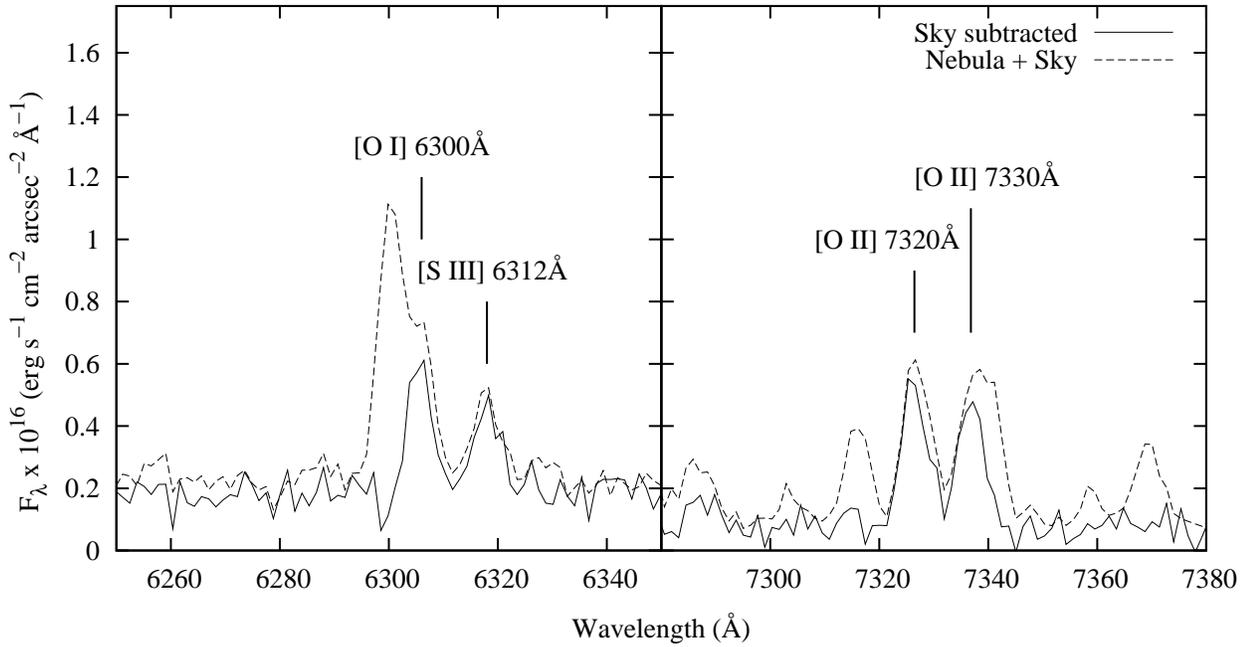}}
\caption{A sample SOAR spectrum near [O I] $\lambda$6300 and [O II] $\lambda\lambda$7320,7330. The solid and dashed lines respectively show the spectrum after and before sky subtraction.}
\label{fig:SoarSpecSample}
\end{figure*}

To remove night sky contamination we obtained a spectrum of the sky 2 deg north and 2.7 deg east of R136. The sky spectrum was smoothed with a median filter spanning 15 arcsec along the slit, to decrease noise and remove stars along the slit. The intensity in the night sky lines was then scaled to match the object frames and subtracted. The redshift of the LMC helps to separate night sky forbidden lines from the same lines from 30 Dor. For example, in the case of the [O I] $\lambda$6300 line, the nebular emission line is redshifted to 6306 \AA. Figure \ref{fig:SoarSpecSample} shows a typical spectrum before and after night sky subtraction, in the regions around 6300 \AA\ and 7330 \AA.

The instrument setup used covers the low ionization [O~I] $\lambda$6300 and [O~II] $\lambda\lambda$7320, 7330 lines as well as the high-ionization [S~III] $\lambda$9068 line and offers a clean unblended measurement of [S~III] $\lambda$6312 redshifted to 6318 \AA. All of these lines except for [S~III] $\lambda$9069 are also within the wavelength range of the Blanco spectra but in that case are heavily blended with night sky lines and/or nebular emission and so were not measured. The region east of R136 was re-observed with SOAR because of the additional constraints which these lines provide when determining the ionization parameter $U$, the energy distribution of the ionizing radiation, and the O/H and S/H abundances.

\section{The spectroscopic 'data cube'}
\subsection{Emission-line flux measurements}

One of our goals was to create a high-quality set of measured emission-line strengths that will be of general public use, so we will describe our measurement procedures in some detail.
Table \ref{tab:LineID} lists the emission lines measured from the Blanco and SOAR spectra. Columns 1 through 3 indicate the observed and rest wavelengths and ionic species. The fourth column lists the extinction coefficient $f_\lambda$ used to deredden the observations, using a standard Galactic extinction curve with a ratio of total to selective extinction R=3.1 ~\citep{Cardelli1989}. The final column indicates which data sets include the emission line.

We extracted 1D spectra binned over successive 2.5 arcsec increments along the slit in the 2D spectra. The 2.5 arcsec width of the extraction window along the slit was chosen so that the in the fainter regions the line flux equaled the  noise in the continuum. The program then automatically measured, in each extracted 1D spectrum, the emission line fluxes and their uncertainty. This produced a `data cube' of emission-line intensity measurements at 4238 points on the 30 Dor nebula.

\begin{figure*}
\center{\includegraphics[width=\textwidth]{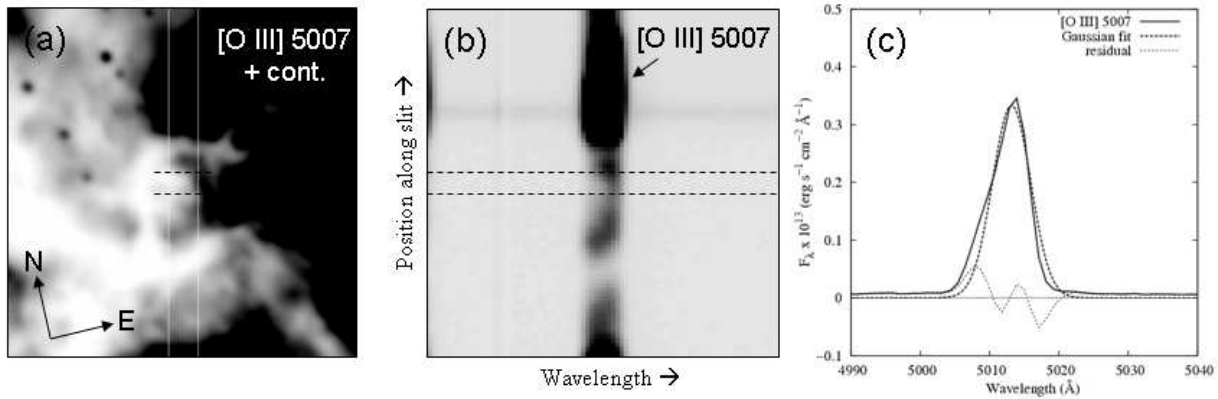}}
\caption{A demonstration that the structure in the line profiles is dominated by spatial structure in the nebula rather than by velocity structure. Panel (a) is a 2D image of the sky in [O~III] emission. Panel (b) shows, on the same spatial scale as panel (a), the 2D Blanco spectrum measured with a 3.5 arcsec-wide slit indicated by the vertical solid lines shown on panel (a). The dispersion is in the horizontal direction. Panel (c) shows the [O~III] $\lambda$5007 profile in an extracted 1D spectrum summed over the portion of the slit between the dotted lines shown in panels (a) and (b).  In panel (c), the top dashed line is the best Gaussian fit and the bottom dashed line is the residual from that fit. The non-Gaussian shape of the [O~III] line in panel (c) clearly is due to the spatial structure seen in panel (a).}
\label{fig:BlancoSpecExample}
\end{figure*}

We needed to use a summation approach to measure the flux in each emission-line profile because of the often complicated shape of the profiles. It is worth noting that most of this structure in the profiles is not due to velocity. It is due to spatial structure in the nebula that is smaller than the 3.5 arcsec slit width used for the Blanco spectra. To illustrate this, Figure \ref{fig:BlancoSpecExample}a shows an enlarged portion of the SOAR [O~III] image at the position of a ring-like structure, and  Figure \ref{fig:BlancoSpecExample}b shows the [O~III] $\lambda$4959 line in the 2D image of the Blanco spectrum crossing this same location. The vertical lines on \ref{fig:BlancoSpecExample}a indicate the region over which the spectrum in Figure \ref{fig:BlancoSpecExample}b was taken. The resulting emission-line structure in Figure \ref{fig:BlancoSpecExample}b  clearly reflects the spatial structure. This resulted in asymmetric profiles which were better measured by summing the total flux. Figure \ref{fig:BlancoSpecExample}c shows the  profile extracted over the area along the slit marked in in the other panels, with a best-fitting Gaussian shape superimposed to demonstrate that the line shape is not well represented by a simple Gaussian.

\begin{figure*}
\center{\includegraphics[width=\textwidth]{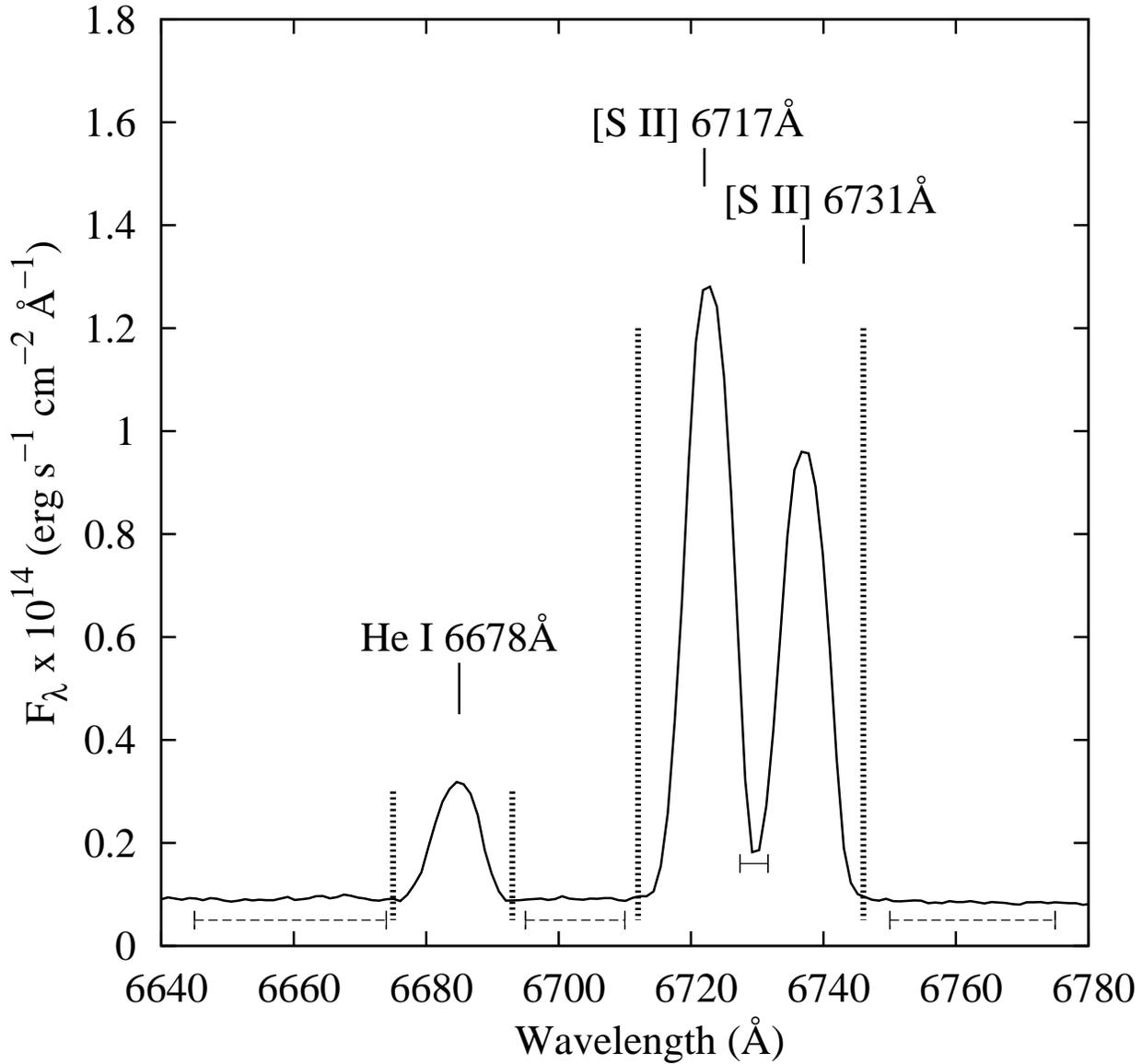}}
\caption{Two examples of extraction windows used to measure line flux. The dashed horizontal bars show the ranges over which the continuum was fitted. The vertical lines show the ranges over which the line fluxes were summed. He I $\lambda$6678 is an example of an isolated emission line, while the [S~II] doublet is slightly blended. The solid horizontal bar between the two [S~II] peaks shows the range over which a search was performed for the minimum value, to define the boundary point between the two individual [S~II] lines, as described in the text.}
\label{fig:FluxWindows}
\end{figure*}

The wavelength windows used in the automated line-measuring procedure were manually set in advance, using the brightest regions of the nebula. Two examples of wavelength windows are presented in Figure \ref{fig:FluxWindows}. In the case of an isolated line like He~I $\lambda$6678, two `wing' boundaries are chosen to include all of the line flux shown by the vertical dashed lines. The continuum regions (indicated by horizontal dashed lines) are separately defined on either side of the emission line, avoiding other emission lines. In the case of mildly blended lines, like the [S~II] $\lambda\lambda$6716, 6731 doublet, the wings are defined to include the total flux from both lines. A search is then made for the minimum value in the area between the two line peaks (over the range indicated by the width of the solid horizontal bar in the example shown in Figure \ref{fig:FluxWindows}), to define the point of separation between the extraction windows of the two individual lines. This does not perfectly deblend the two lines, but since the ratio of [S~II] $\lambda$6716/ $\lambda$6731 is always of order unity the error is negligible.

This procedure for measuring the line strengths works well even in the cases where stars are present in the extraction region so long as the star is not extremely bright or the nebula extremely faint. Possible problem cases were identified using the continuum bins, and were checked manually and for a few cases all emission lines at that position were excluded. In the case of bright stars with strong absorption features the surface brightness of \Hb~ will be underestimated. Placement of bright star clusters in the slit was intentionally avoided and the number of regions affected by stellar absorption is small.

\subsection{Noise estimates}
We estimated the uncertainties in the measured line strengths by assuming Poisson statistics in the original photon count rate and then carrying parallel noise images all the way through the same reduction process as the data images. At the beginning of the data reduction the measured counts were converted to photons using the known detector gain. An initial noise image was then calculated using

\begin{equation}
\sigma = (N_{photons} + gain \times ReadNoise^{2}_{electrons})^{1/2}
\end{equation}. 

All multiplicative calibrations such as flat fields, illumination corrections and the flux calibration were applied to the noise image by multiplying the $\sigma$ values in the noise image.

In the step where the spectra were linearized and distortion was corrected, there was a re-binning of the data. In this case $\sigma^2$ (the variance of the noise) was rebinned using the same transformation. The resulting $\sigma$ in each pixel is thus summed in quadrature. We manually inspected a random set of extracted spectra and verified that these estimated errors do in fact reasonably measure the observed pixel-to-pixel scatter in the continuum points. An exception to this is [O~III] $\lambda$4363, where uncertainties in the continuum level and blending with the wings of H I $\lambda$4340 cause special problems, as is discussed below. The $\sigma$ values for each pixel were then used to compute error bars for the measured line fluxes and ratios.

\subsection{Detector saturation}
The large range in the surface brightness of the nebula meant that the 5 arcmin long slit included both bright and faint regions. In order to get a sufficiently high signal/noise ratio in the faint regions, long exposures were used and the [O~III] $\lambda$5007 and/or \Ha~ lines were often saturated at various places along the slit. When necessary, additional short exposures were made to correct for saturation. All repeated observations of a single position, long and short, were gray-shifted to the highest value according to the total flux of the [O~III] $\lambda$4959 line summed over the full 300 arcsec length of the slit. The $\lambda$4959 line was used because it is a moderately strong line but is never saturated. Summing along the slit minimizes any effects of variable seeing or telescope drift.

\begin{figure*}[]
\center{\includegraphics[width=\textwidth]{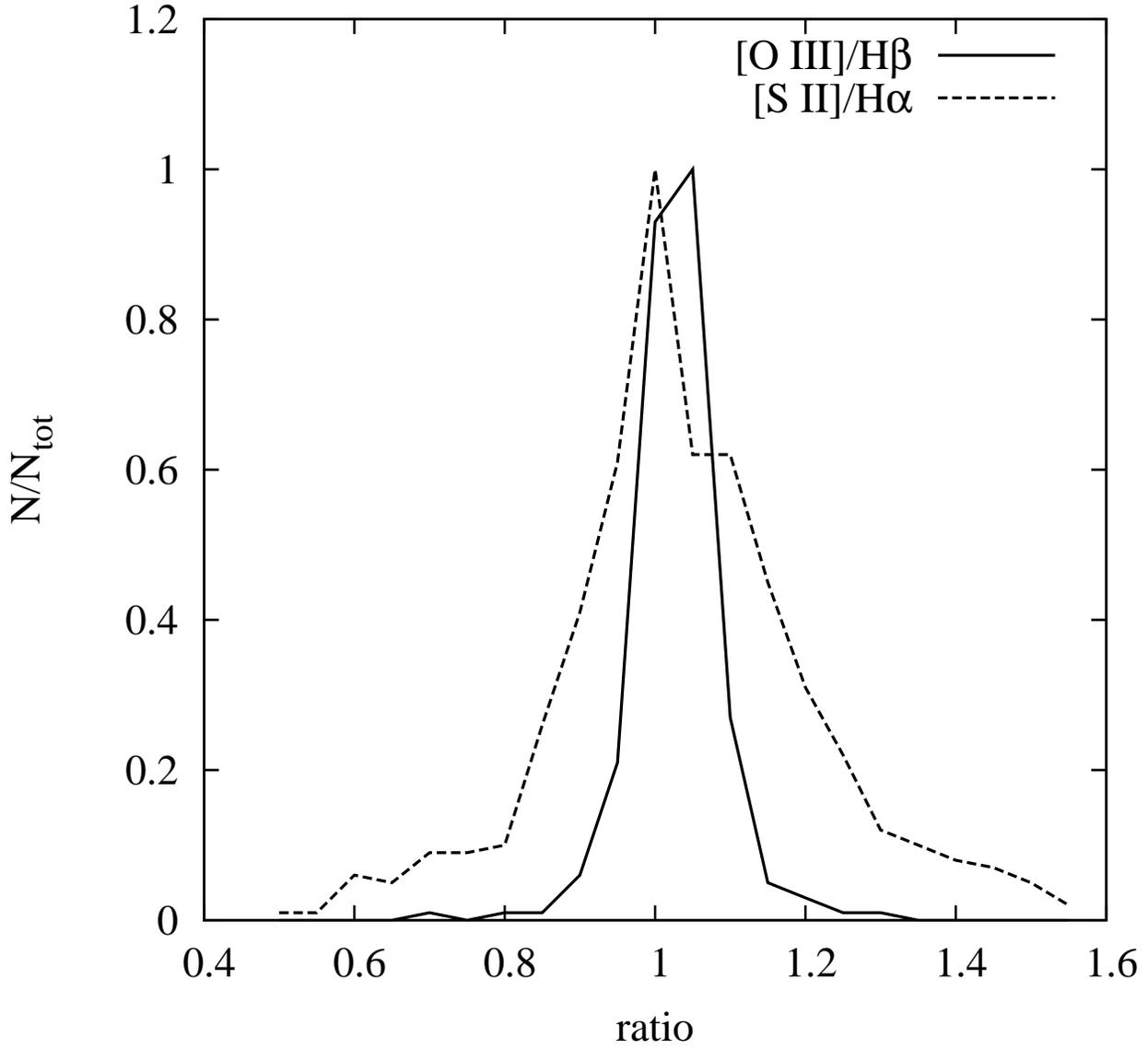}}
\caption{Repeatability of results at overlapping points along different Blanco slit positions. The curves are histograms of the distributions of ratios of intensity ratios; for example the [OIII]/\Hb~ measured from one slit position is divided by the [OIII]/\Hb~ measure at the same position on the sky but from a different slit position.}
\label{fig:BlancoRepeatHist}
\end{figure*}

As a final check to the data, line ratios of extracted regions were compared wherever two slits intersected.  While the extracted regions at these points overlap, they do not sample completely identical parts of the nebula. The normalized histogram shown in Figure \ref{fig:BlancoRepeatHist} demonstrates the repeatability of the [O~III] $\lambda$5007/\Hb~ and ([S~II] $\lambda$6716+$\lambda$6731)/\Ha~ ratios measured for these overlapping observations. The histogram shows that the data do repeat to within the typical range of differences due to the mismatch between the areas sampled on the sky by the two slits, with especially good repeatability in the [O~III]/\Hb~ ratio.

\subsection{Reddening correction}
The reddening was determined separately for each extracted 1D spectrum, using the observed \Ha/\Hb~ intensity ratio and assuming an intrinsic ratio of 2.87 appropriate for Case B and a gas temperature $10^4$ K ~\citep{agn2006} and R=3.1. This is adequate even though 30 Dor is in the LMC because (1) although the LMC extinction curve departs considerably from the Galactic curve in the ultraviolet, the two are very similar in the optical passband, and (2) some considerable part of the reddening is in any case due to foreground material within our own Galaxy. We then applied the reddening to each measured line from the same extracted 1D spectrum, using the extinction curve values $f_\lambda$ from Table \ref{tab:LineID}. A check on the validity of this procedure is that the dereddened H$\gamma$/\Hb~ ratios agree with the predicted Case B values to within an average of 3 percent, with a 1$\sigma$ scatter of 4 percent which is consistent with the observational errors.

\subsection{Electron density and temperature}
At each position along  each spectrograph slit we determined the electron density $n_e$ from the dereddened [S~II] $\lambda$6716/$\lambda$6731 line ratio. The IRAF/STSDAS task $nebular$ was used to derive the electron densities using a 5 level sulfur atom assuming a gas temperature of $10^4$ K. The errors in the density measurements were calculated using the 1$\sigma$ uncertainty in the line ratio. The difference between the densities at these extremes and at the nominal value are the reported density uncertainties. If either the nominal or 1$\sigma$ [S~II] line ratio was greater than 1.41 the density was only constrained to be $n_e \le 10$ cm$^{-3}$ (the low-density limit). Using the published contour plots of log($ n_e$) derived from [S~III] in I09 we compared the electron densities measured from [S~II] and [S~III]. We find they agree to 0.1 to 0.2 in log($ n_e$), noting that the increments in the density contours of I09 are 0.1 in log($ n_e$).

The electron kinetic temperature $T_e$ was also measured at each position, from the dereddened [O~III] ($\lambda$5007+$\lambda$4959) / $\lambda$4363 ratio again using the IRAF/STSDAS nebular task. The density used in the calculation is from the [S~II] lines. Errors were computed in the same way as for the density values.

\subsection{A publicly available data set}
The Blanco instrument setup was chosen to cover the strong optical nebular emission lines from 4100 to 7400 \AA\ including \Hb, \Ha~ and the [S~II] doublet at $\lambda\lambda$6716, 6731. The Blanco observations were made in the form of a grid with fairly regular spacing which allow the data to be turned into a `data cube'.

Our Blanco slit positions were chosen to produce a relatively unbiased sample of the diffuse emission as well as to include discrete structures of interest. Post-observing verification of the location of each slit position on the sky was done using narrow band emission line and continuum images taken at the SOAR telescope by (1) comparing the extracted surface brightness profile of \Ha~ along the slit to the \Ha~ images, (2) using positions, measured from our narrow band continuum images, of stars purposely placed in the slit as references and (3) where necessary using 2D nebular features visible in both the spectra and images such as in the example shown in Figure \ref{fig:BlancoSpecExample}.

Table \ref{tab:BlancoFluxes} is a sample of the final data product from the Blanco spectra. It lists the dereddened line strengths and several additional parameters for each extraction window along each slit position, and their uncertainties. The entire table, which is available in electronic format, includes measurements at 4238 positions with one row per position. Column 1 lists the slit position number (as used in Figure \ref{fig:HaMosaic}b
and Table \ref{tab:SPECpointings}). Columns 2-4 contain the RA and Dec offsets in arcsec from R136 and the central pixel row of the extraction window along the slit. Then columns 5-10 list the electron density and its 1$\sigma$ uncertainty limits, followed by the electron temperature with its 1$\sigma$ uncertainties. Column 11 is the reddening $A_V$ deduced from the \Ha/\Hb~  ratio. These are followed in columns 12 and 13 by the dereddened \Hb~ surface brightness and its 1$\sigma$ uncertainty. The remaining 44 columns list, in pairs, the dereddened surface brightness of each measured emission line relative to \Hb~ and the corresponding uncertainty.

The last three rows of Table \ref{tab:BlancoFluxes}, collectively labeled position 38, list the average properties of 30 Doradus derived from the entire data set. Position 38 row 1 is the average of the dereddened fluxes, and the physical properties derived from those. Row 2 is the average of the observed values. Row 2 was then dereddened using the average \Hb~ and \Ha~ fluxes to produce row 3. Rows 2 and 3 represent the global spectroscopic properties that would be measured for 30 Dor if it were spatially unresolved.

Table \ref{tab:SOARFluxes} is the similar data product from the SOAR spectra, computed in exactly the same way as for Table \ref{tab:BlancoFluxes}. The SOAR spectra are presented in a separate table because there are a different number of columns due to the additional emission lines which were measured, including [S~III] $\lambda$9069. The ratio [S~III] $\lambda$9069/$\lambda$6312 is sensitive to the gas temperature like [O~III] $\lambda$5007 / $\lambda$4363. The temperatures measured from the [S~III] lines are included in Table \ref{tab:SOARFluxes}.

In addition to the tabulated data the fully reduced Blanco 2D spectra, noise and calibration frames are publicly available.  They are posted on a permanent web site at the Michigan State University Department of Physics \& Astronomy\footnote{http://www.pa.msu.edu/astro/thesis/pellegrini/30dor/}. The complete version of Tables \ref{tab:BlancoFluxes} and \ref{tab:SOARFluxes} are available in electronic form at that site, and also on the ApJS website.

\section{Observational Results }
\subsection{Overview}
\begin{figure*}
\includegraphics[scale=0.75]{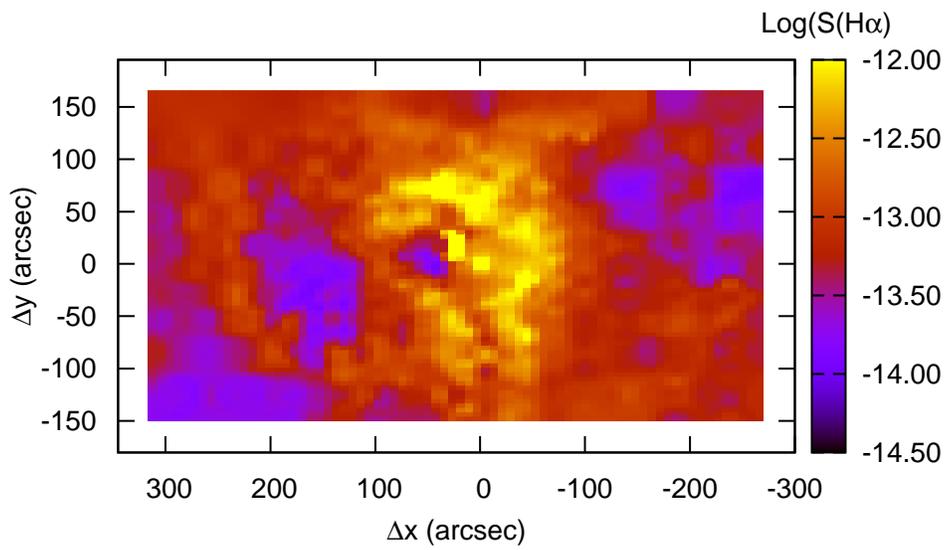}
\caption{Interpolated dereddened \Ha~ surface brightness in erg s$^{-1}$  cm$^{-2}$ arcsec$^{-2}$. The region shown is the same as that outlined in Figure ~\ref{fig:HaMosaic} and Figures \ref{fig:MapAv} through \ref{fig:MapS2}. R136 is at $\Delta x$ = 0, $\Delta y$ = 0. }
\label{fig:MapHa}
\end{figure*}

\begin{figure*}
\includegraphics[scale=0.75]{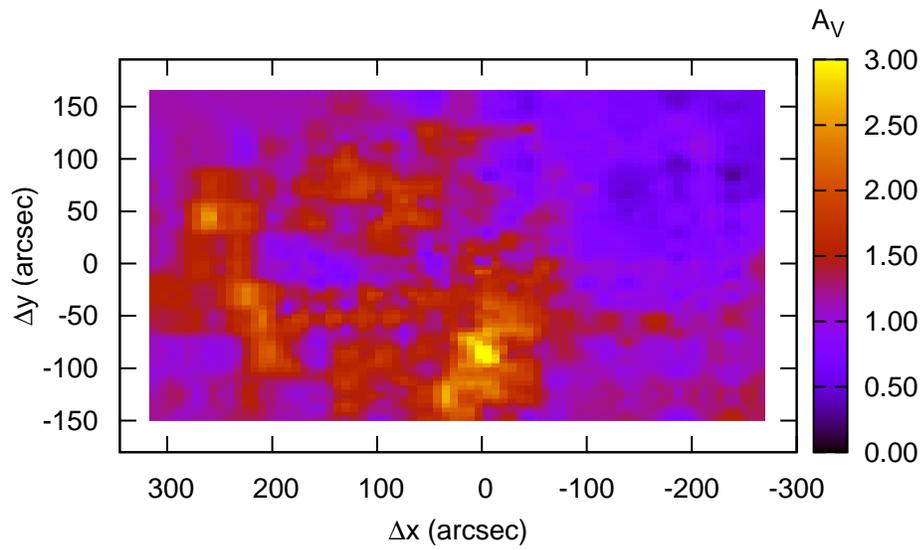}
\caption{Interpolated $A_V$.}
\label{fig:MapAv}
\end{figure*}

\begin{figure*}
\includegraphics[scale=0.750]{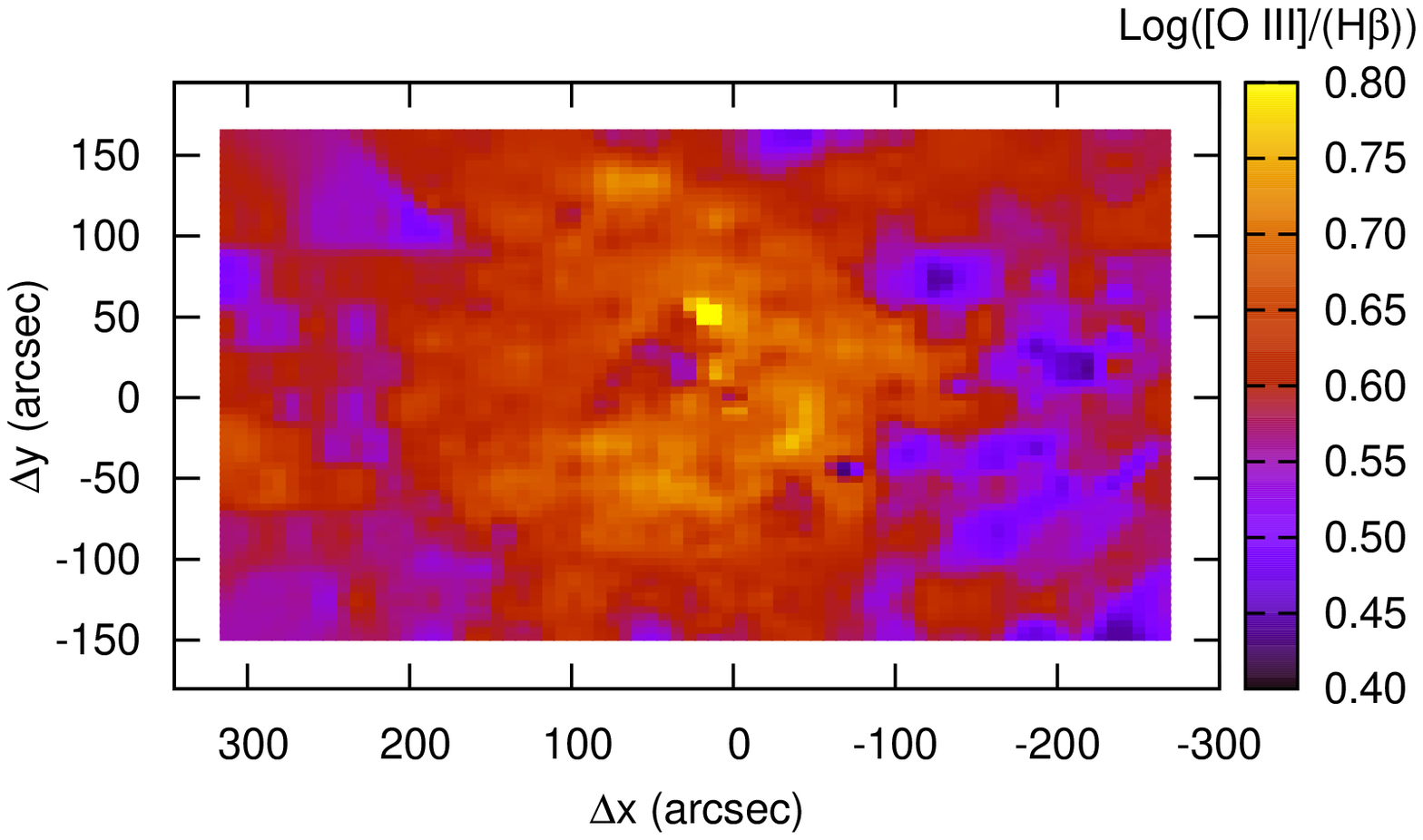}
\caption{Log([O~III]/\Hb).}
\label{fig:MapO3}
\end{figure*}

\begin{figure*}
\includegraphics[scale=0.75]{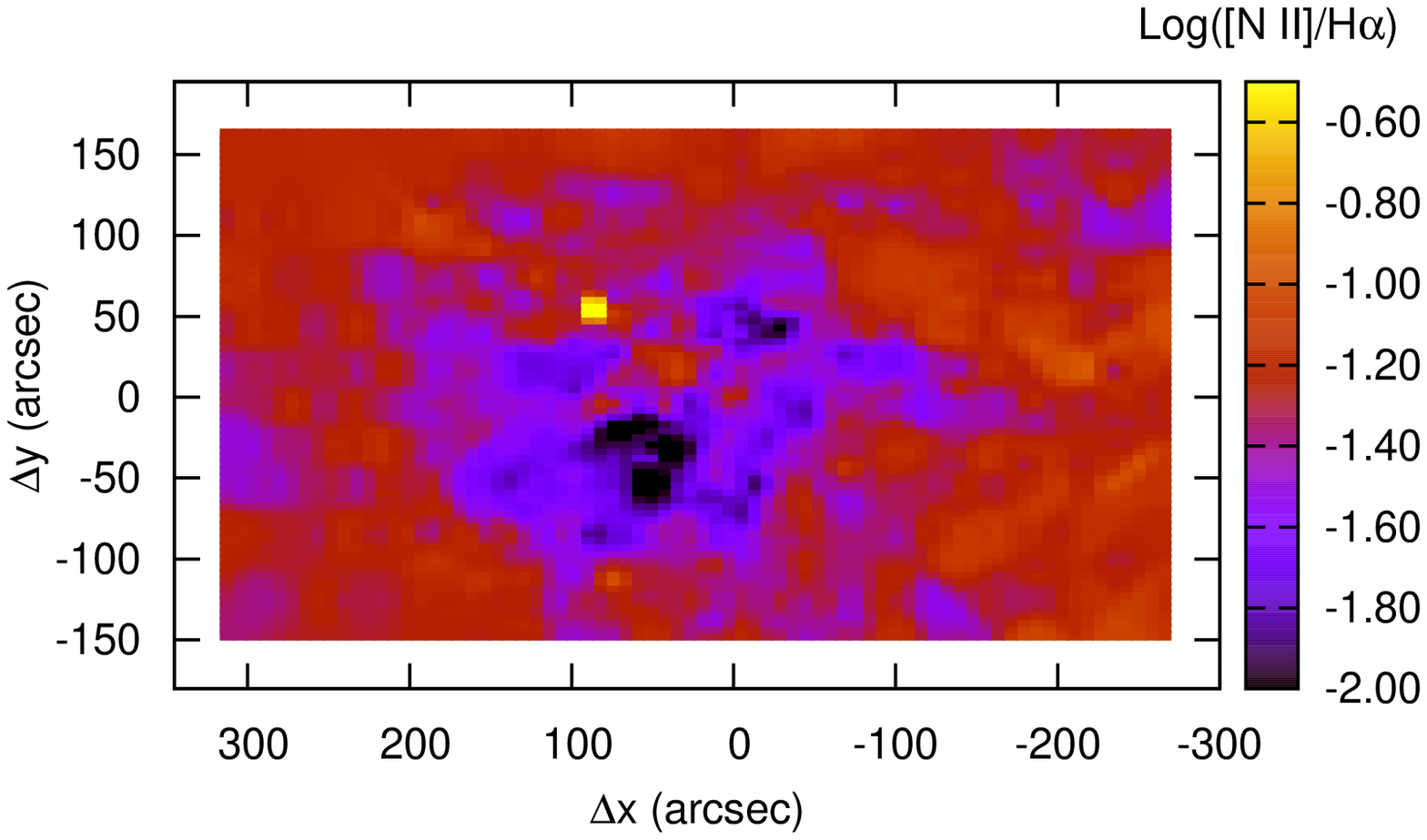}
\caption{Log(([N~II] $\lambda$6584)/\Ha).}
\label{fig:MapN2}
\end{figure*}

\begin{figure*}
\includegraphics[scale=0.75]{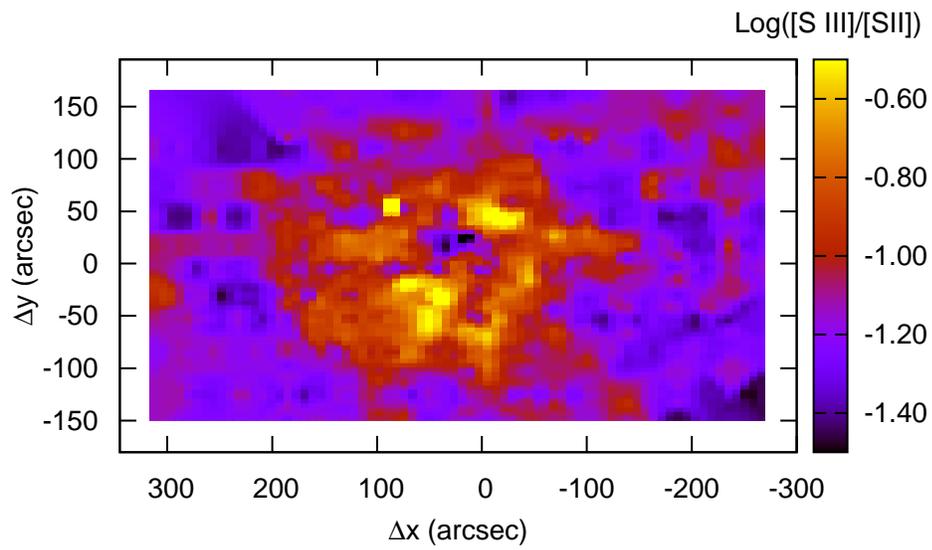}
\caption{Log([S~III] $\lambda$6312 / ([S~II] $\lambda$6716+$\lambda$6731)).}
\label{fig:MapS23}
\end{figure*}

\begin{figure*}
\includegraphics[scale=0.75]{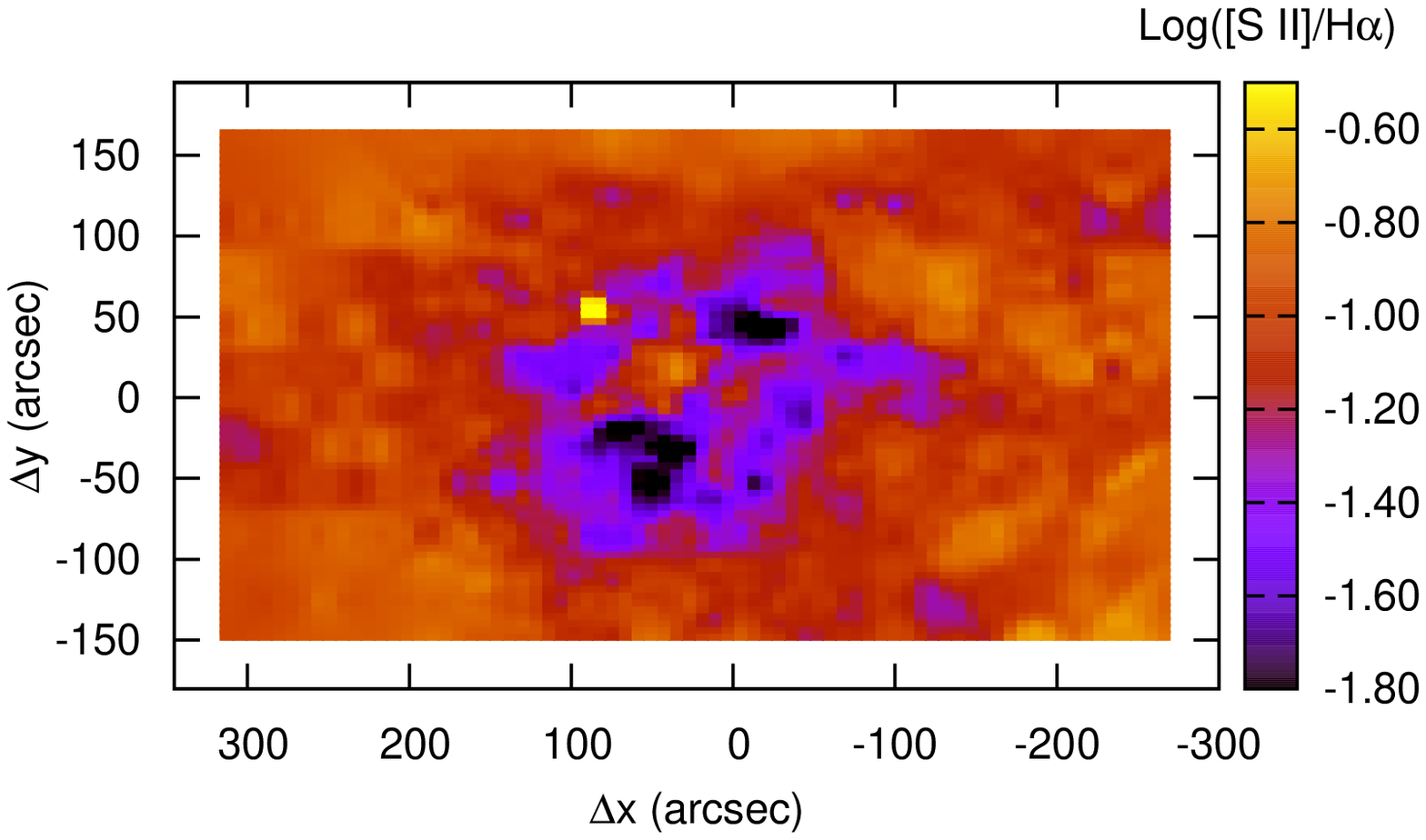}
\caption{Log(([S~II] $\lambda$6716+$\lambda$6731)/\Ha).}
\label{fig:MapS2}
\end{figure*}

Figures \ref{fig:MapHa}-\ref{fig:MapS2} are maps of selected emission lines, nebular diagnostics, and physical conditions derived from those emission lines, interpolated in the spatial plane of the Blanco data cube. The two extreme slit positions, 16 and 17, are excluded from these maps.  These maps are rotated 13 deg with respect to the N-S and E-W directions, and are labeled with offsets in that rotated coordinate system. An outline of the area covered by these maps is shown on Figure \ref{fig:S2Ha_ratio}.

Figure \ref{fig:MapHa} shows the dereddened \Ha~ surface brightness and demonstrates the effective resolution of the grid interpolated from the 'data cube'. It can be compared with the SOAR \Ha~ image in Figure \ref{fig:HaMosaic}. Most features, including the bright arcs centered around NGC 2070, are visible. To the east the cavity-like region of low surface brightness as well as the bright rim on its eastern edge are clearly seen.


The ionization of the gas is traced by intensity ratios which include  ([O~III] $\lambda$5007)/\Hb~  (Figure \ref{fig:MapO3}),  ([N~II] $\lambda$6584)/\Ha~ (Figure \ref{fig:MapN2}),  ([S~III] $\lambda$6312)/([S~II] $\lambda$6716 + $\lambda$6731) (Figure \ref{fig:MapS23}), and [S~II]/\Ha~ (Figure \ref{fig:MapS2}). Figure \ref{fig:MapS2} shows the ratio derived from the Blanco spectra, as opposed to the ratio made from the SOAR images that is shown in Figure \ref{fig:S2Ha_ratio}.  These line ratios consistently show a rough circular symmetry around a point about 40 arcsec E of R136. The same distribution was identified by I09 from a Spitzer Space Telescope survey of 30 Dor using higher ionization lines. This will be discussed further below.


\subsection{[O~III] kinetic temperature}
The electron kinetic temperature is computed from the [O~III] $\lambda$4959 + $\lambda$5007 / $\lambda$4363 ratio.  We find the nebula to be fairly isothermal within a few thousand degrees, in the range 9,000 $\le T_e \le$ 12,000 K, measured using the data points with nominal error bars smaller than 10 percent.

The accuracy of the overall scale for the gas temperature will be an important point in our analysis in Paper II. Here we compare our mean [O~III] temperature to that found by ~\citet{KrabbeCopetti2002}, who used three long-slit spectra which criss-crossed the central region of 30 Dor with much of their length falling on the bright arcs. There are a number of valid ways to calculate an average temperature $\langle T_e \rangle$ for 30 Dor. For the most direct comparison to distant, unresolved GEHRs, we should spatially integrate, over the whole nebula, all of the light in each emission line, and then apply a single reddening correction based on the ratio of the spatially integrated \Ha~ and \Hb~ fluxes. From those reddening-corrected line strengths we can then find $\langle T_e \rangle$ using the observed [O~III] ($\lambda$4959 + $\lambda$5007) / $\lambda$4363 ratio $R_{OIII}$ given by

\begin{equation}
\label{eq:RO31}
R_{OIII} = \frac{\int{F(4959\AA + 5007\AA)d\Omega}}{\int{F(4363\AA)d\Omega}}
\end{equation}. 
This method uses the total line fluxes listed for Position 38, row 3 in Table \ref{tab:BlancoFluxes}, and yields an equivalent temperature of  $\langle T_e \rangle$ = 10,680$\pm5$ K ($R_{OIII}$ = 172$\pm0.2$). Here we have used the error bars on the measured line strengths derived as explained in the previous section.

An alternative is to use [O~III] line fluxes that have first been individually dereddened at each point on the nebula and then spatially integrated. That method leads to $\langle T_e \rangle$ = 10,760$\pm8$ K ($R_{OIII}$ = 168$\pm0.4$). This corresponds to the values listed for Position 38, row 1 in Table \ref{tab:BlancoFluxes}.  

When individual measurements of $T_e$ have been made at multiple locations in a nebula, as is the case here, another common approach is to compute
\begin{equation}
\label{eq:RO32}
\langle T_e \rangle  = \frac{ \int{ T_{e}(\vec{r}) \times F(H\beta) d\Omega } }{ \int{F(H\beta) d\Omega}}
\end{equation} 

\noindent where $T_e$ is weighted by the \Hb~ flux at each position. Using this method we again find $\langle T_e \rangle$ = 10,680 K, equivalent to $R_{OIII}$ = 172.

For comparison, using this last method, ~\citet{KrabbeCopetti2002} found $\langle T_e \rangle$ = 10,270 K ($R_{OIII}$ = 195) using measurements along all three of their slit positions. An even lower $\langle T_e \rangle$ = 9,990 K  ($R_{OIII}$ = 209) was found by ~\citet{Tsamis2003} who summed over the length of a 160 arcsec slit. Even though the formal standard error of the mean of each of these mean temperatures is $\sim$10 K, the true uncertainties will be somewhat larger.

To explore the source of the discrepancy between our results and the previous measurements of $\langle T_e \rangle$, we coadded all of our Blanco spectra to produce a single spectrum with very high signal/noise ratio, excluding only the regions containing the brightest stars. The entire spectrum was then dereddened as in the first of the averaging methods described above. We then measured $R_{OIII}$ and computed the associated $\langle T_e \rangle$ in several ways. 

First, using our automated line measuring software on this spectrum produced $\langle T_e \rangle$ = 10,565 K ($R_{OIII}$  = 178), in reasonable agreement with the value  $\langle T_e \rangle$ = 10,680 K obtained from the first of the averaging methods described above.

We then investigated the way in which our program sets the continuum level. The automated software fits the continuum in windows covering the wavelength ranges $\lambda\lambda$4280-4325 \AA\ (blueward of \Hb) and $\lambda\lambda$ 4410-4150 \AA\  (redward of [O~III] $\lambda$4363). A manual measurement of the [O~III] line strengths using the IRAF splot routine and similar continuum levels produced essentially the same $R_{OIII}$   and  $\langle T_e \rangle$, verifying that our automated routine works correctly.  However, inspection of the coadded spectrum showed that the region between \Hb~ and [O~III] $\lambda$4363 is partly filled in by a faint plateau due to line wings or weak emission, and that the continuum also has complications in the region around [O~III] $\lambda$4959 and $\lambda$5007.  We tried various choices of how to draw in the continuum and were able to measure line strengths corresponding to a temperature as low as $\langle T_e \rangle$ = 10,280K ($R_{OIII}$ = 194.6).

We conclude that the value $\langle T_e \rangle$ = 10,270 K found by ~\citet{KrabbeCopetti2002} agrees with our corresponding value $\langle T_e \rangle$ = 10,673 K to within the measurement uncertainties due just to the issue of where the continuum is drawn. This uncertainty in the measured temperature translates to an uncertainty in the derived oxygen abundance. There would be a 12 percent (0.05 dex) increase in the derived O/H abundance ratio if the temperature were decreased from 10,680 K to 10,280 K. The $\lambda$5007/$\lambda$4363 ratio measurements at individual points show a much larger scatter, with a standard deviation $\sigma$ = 50 per cent (0.29 in the log). Although this scatter is much larger than the uncertainty computed at each point from the photon counting statistics, we interpret it as being dominated by the true point-to-point measurement uncertainties.

We cannot resolve the discrepancy between the temperatures measured by ~\citet{Tsamis2003} and those measured here and by ~\citet{KrabbeCopetti2002}. However, we are confident that our measurements are correct to within the uncertainties discussed above.

\subsection{Ionization mechanism}
The observed kinetic temperature provides an important constraint to the energy sources that ionize the gas. The violent history of 30 Doradus is evident in the diffuse X-ray emission that is present throughout the nebula ~\citep{Townsley2006}, as well as in the high-velocity expansion features with  speeds up to 200 km s$^{-1}$ seen in \Ha~ emission ~\citep{ChuKennicutt1994}. These show that supernovae and strong winds from massive O and WR stars have combined to heat gas to the observed 3.5 -- 7$\times10^6$ K temperatures derived from Chandra X-ray spectra. Despite these indications of high-velocity flows, it has long been known that the moderate $10^{4}$ K temperatures in the ionized gas do not indicate shock heating as the current source of ionization.

The absence of [O~IV] or [Ar~V] emission lines (I09) as well as of any significant detection of nebular He II emission has been used to argue that strong shocks are not an important ionization mechanism in 30 Doradus. Instead, photoionization is implied as the energy input mechanism. However, this still leaves open the question of the source of the ionizing photons, which we now consider.

\section{Structural Details}
The geometry and ionization structure of 30 Dor are quite complex. The SOAR direct images provide a powerful tool for distinguishing edge-on ionization fronts (IFs), regions ionized by localized sources of radiation besides the central cluster, optically thick pillars like those found in M16 (Hester 1996), and line-emitting foreground structures. In particular, the nearly reddening-free [S~II]/\Ha~ image (Figure \ref{fig:S2Ha_ratio}), formed from the ratio of the SOAR images in those two passbands, offers considerable insight into the ionization structure on many scales.

\subsection{Embedded ionizing stars}
A key question is the degree to which the central cluster dominates the ionizing radiation field at different positions in the nebula. This will depend on the extent to which isolated O stars provide additional local contributions to the radiation field. Such stars can be recognized by localized, circular variations in the ionization level as indicated in the [S~II]/\Ha~ maps, in combination with brightness enhancements in the \Ha~ emission. We carefully searched the SOAR images for such features, and located the 49 isolated stars listed in Table \ref{tab:LocalSources}. The spectral types of known stars within the affected region are listed. Sources identified in previous surveys with with unknown spectral types are listed as unknown and previously unidentified objects are left empty. Of the 49 sources identified, 19 have stars with known spectral types. There are 4 WR/WN, 7 O and 6 B type stars. The later O and early B type stars produce less ionizing radiation than more massive stars and may be illuminating some of the remnant material from their formation still present due to weak winds. The other objects identified with unknown spectral types are typically IR sources. Given the fraction of massive stars identified as embedded objects, the unknown sources should be followed up to determine if they too are massive stars. Some percentage of these is likely to have been selected by chance, so this estimate should represent an upper limit to the number of locally ionizing stars. The area of the nebula in which the ionization level is significantly affected by the radiation field of these stars, as projected on the sky, covers only 2 per cent of the full nebula. This means that it will not be an important contaminant in our analysis, in Paper II, of the overall properties of the nebula.

\subsection{Dense pillars}
There are numerous bright fingers and protruding IFs scattered over the face of the nebula, but they are particularly concentrated in the large X-ray emitting cavity to the East of R136. Many of these are probably elephant trunks similar to the famous ``Pillars of Creation'' in M16. A small number of these have been commented on by ~\citet{Scowen1998}. These structures are easily identified using a combination of the \Ha, [S~II] and 8$\micron$ images.  Table \ref{tab:Pillars} catalogs 106 of the brightest such features, listing their RA and Dec positions, projected length and PA. However, there are hundreds more that blend together as one looks on finer and finer scales, until they are indistinguishable from small examples of the edge-on ionization fronts discussed above. Almost all of these point back to R136, strongly suggesting that the central cluster is the dominant source of ionizing radiation.

\begin{figure*}
\center{\includegraphics[scale = 5]{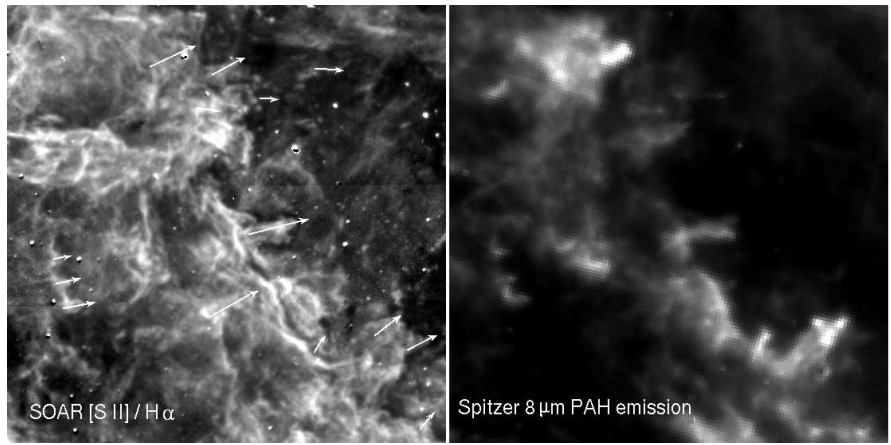}}
\caption{Left SOAR [S~II]/\Ha; Right SPITZER 8$\micron$ PAH. A selection of bright pillars are shown with arrows indicating their location and direction. These dense IFs are detected in both optical and IR passbands and show a connection with the background molecular cloud.}
\label{fig:Pillars}
\end{figure*}

Many of these features are clearly connected to larger bodies of molecular gas, as can be seen by comparing our [S~II]/\Ha~ ratio image to the 8 $\micron$ image from the Spitzer Space Telescope archives, which shows PAH emission (Figure \ref{fig:Pillars}).  However, some of them do appear to be isolated tubes or blobs of gas rather than protrusions from background walls of molecular material. ~\citet{Scowen1998} have commented on one bright structure of this latter type. Some of these might correspond to the expanding shells seen in \Ha~ emission in the high-resolution echelle spectra of ~\citet{ChuKennicutt1994}.

\subsection{Edge-on ionization fronts and the source of the ionizing radiation}
\begin{figure*}
\center{\includegraphics[scale=0.80]{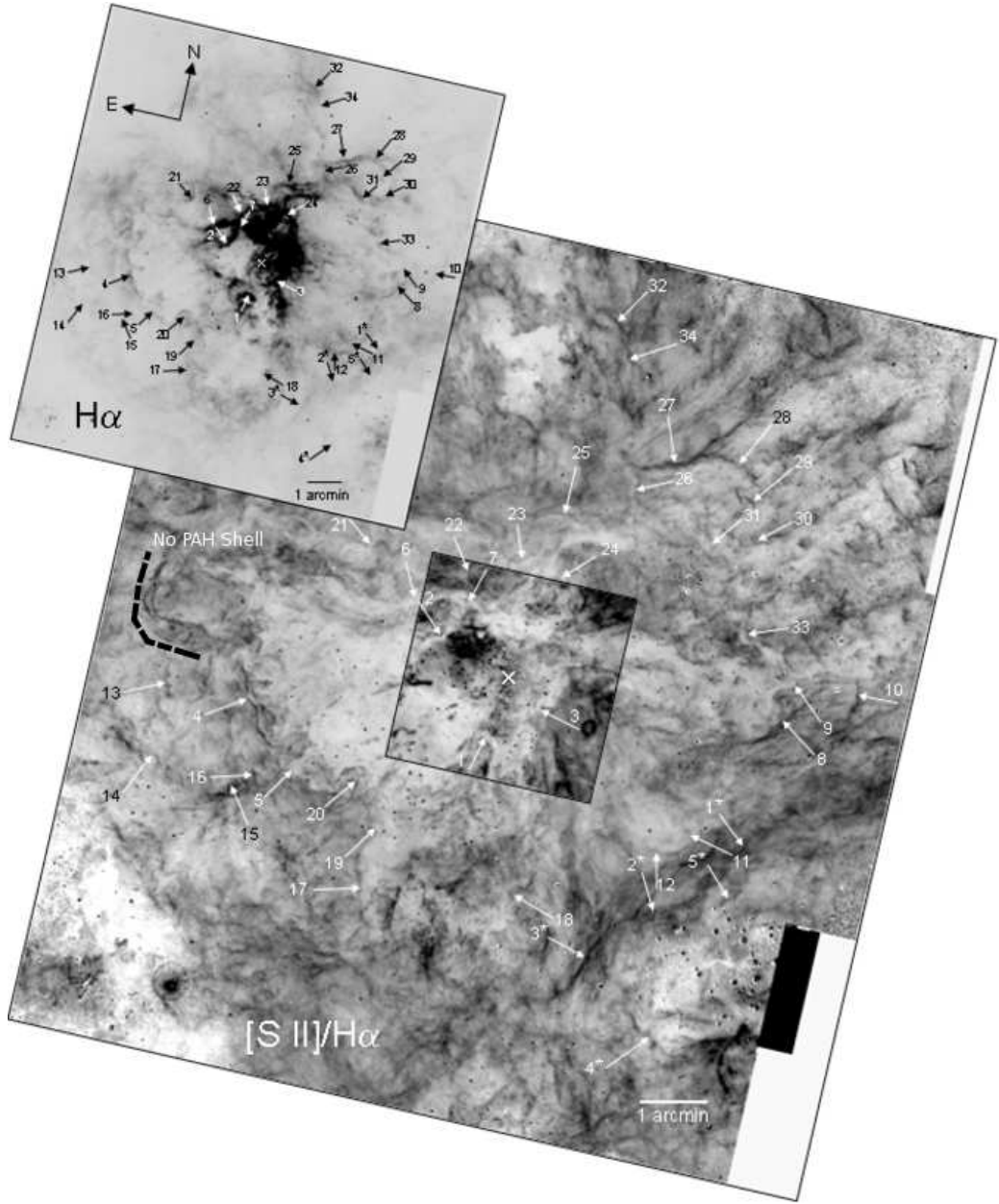}}
\caption{\Ha (upper) and [S~II]/\Ha~  (lower) images showing the positions of the ionization fronts (IFs) catalogued in Table 8. The approximate center of each IF is at the tip of the arrow marking it. The arrows point in the direction towards higher ionization within each I-front, and except for IF16 are approximately perpendicular to the IF. On the main image, darker shades indicate larger values of [S~II] /\Ha, but with the grey scale adjusted in the central box to show regions with lower [S II] /\Ha~ ratios. The white X in each image marks the position of R136.}
\label{fig:IFsFinder}
\end{figure*}

Edge-on ionization fronts stand out in the [S~II]/\Ha~ image as narrow linear structures with high [S~II]/\Ha~ ratio, butted up against regions of low [S~II]/\Ha. 
Table \ref{tab:IFs} lists prominent, isolated, edge-on ionization fronts found both in our [S~II]/\Ha~ image and in the Spitzer 8 $\micron$ PAH mosaic. The structures identified in both the optical and IR passbands are perfectly suited for detailed, high angular resolution (e.g. with HST and/or ALMA) studies of the neutral and molecular gas beyond the IF, commonly called the Photo-Dissociation Region or Photon Dominated Region (PDR). A finding chart of all IFs listed in Table \ref{tab:IFs} is shown on a [S~II]/\Ha~ image in Figure \ref{fig:IFsFinder}. 

With our new observations there is enough information available about several of these IFs to test whether or not they are likely to be photoionized by the central cluster including R136. The thickness of the ionized gas layer $dr$, depends on the density $n$ and the distance of the gas from the source of ionizing radiation $r_0$ according to

\begin{equation}
\label{eq:dr}
dr\propto \frac{Q_0}{4 \pi \alpha n_e n_p r_{0}^{2}}
\end{equation}
where $Q_0$ is the ionizing photon luminosity by number, $\alpha = 2.59\times 10^{-13}$ cm$^{-3}$ s$^{-1}$ is the H$^{+}$ recombination rate, and we assume $dr \ll r­_0$. The test consists of computing $Q_0$ values for different IFs at very different distances from R136, assuming that R136 marks the center of the ionizing radiation, and seeing if the different $Q_0$ measurements agree with each other. There will be considerable uncertainty due to the implicit assumption that the projected distance from R136 to each IF is the true three-dimensional distance, so we do not carry error bars through the analysis and will regard agreement at a factor of two level to be satisfactory. 

Two IFs of interest for this test, IF1 and IF2, are fairly close to 30 Dor (see Figure \ref{fig:IFsFinder}). Both have a spatially resolved layer of H$^+$ separated from an edge-on IF seen in [S~II], similar to the Orion Bar (see P09).

In the case of IF1, there is a line of stars coincident with the IF that could be an alternative to NGC 2070 as the source of ionization, or which could be stars formed in a region of gas that was compressed by radiation and wind pressure from the cluster, or which could simply be foreground stars. Most of these stars fell in one or another of our slit positions. To gauge the influence these stars may have on the ionization structure of this region, we have examined the stellar spectra and found them to lack any absorption features. This would indicate they are massive stars; however this result is uncertain due to the brightness of the emission lines coincident with the expected absorption features. If these stars were embedded in the gas we would expect the stellar spectra to be significantly reddened, but with the possible exception of one star, this does not appear to be the case. Additionally, we note that the overall structure of IF1 is oriented almost exactly perpendicular to the direction to R136. There are also many prominent pillars in this region, all pointing back up to R136, suggesting the nearby stars are unimportant to the structure.

\begin{figure*}
\center{\includegraphics[scale=0.5]{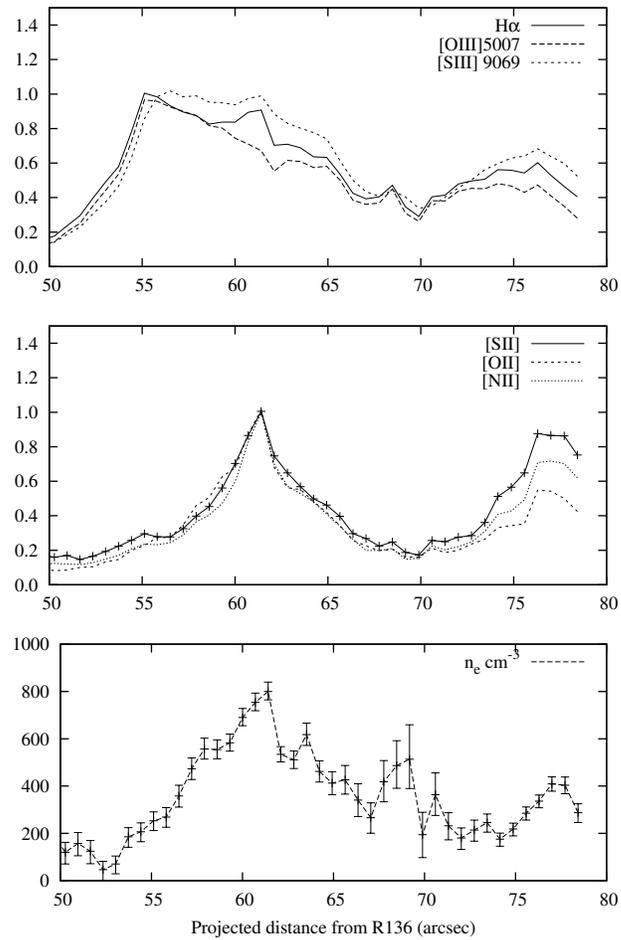}}
\caption{The profile of IF1. Top: Ionized gas is traced by \Ha, [OIII] and [S~III] emission. Middle: The IF is traced by [S~II], [N~II] and [O II]. Bottom: The electron density profile in cm$^{-3}$ measured from [S~II] as described in the text.}
\label{fig:IF1Profile}
\end{figure*}

Our SOAR spectrographic slit position number 40 cuts directly across IF1, avoiding the pillars noted by ~\citet{Scowen1998} and targeting the larger structure. The emission-line strengths in the region of IF1 were measured as described above, with the exception that the extraction window is smaller (0.75 arcsec) along the slit to better resolve the detailed structure. Figure \ref{fig:IF1Profile} shows the clear separation of the highly ionized gas from the lower-ionization gas in IF1. In the top panel are the relative intensities of \Ha, [O~III] $\lambda$5007 and [S~III] $\lambda$9069 which trace highly ionized gas. The middle panel shows the same for [S~II], [O~II] and [N~II], all tracers of the IF. The bottom panel shows the linear density profile of the structure. At the illuminated face the structure has an electron density $n_e \simeq 100$ cm$^{-3}$, increasing to 800 cm$^{-3}$ at the peak of the [S~II] emission, and then decreasing to a minimum of 200 cm$^{-3}$.

There is a bright knot in the [S~II] images at the location of the peak density. Excluding this point, the density at the IF is likely to be 560 cm$^{-3}$. Using this density, assuming a central ionizing source, and that the 13.6 pc (58 arcsec) projected distance from R136 is the true three-dimensional distance and that the 1.18 pc projected thickness of the $H^+$ zone is the true thickness, Eq. \ref{eq:dr} requires $Q_0 = 5.9\times10^{51}$ s$^{-1}$.  In addition to the uncertainties in projected vs. true distances, this value neglects the effect the observed gradient will have in lowering the measured average $n_e$.

The edge-on front IF2 forms the inner edge of one of the most prominent bright arcs, located at a projected distance of 17 pc (70 arcsec) NE of R136. The thickness of the H$^+$ layer perpendicular to the direction to R136 is approximately 0.94 pc. The density at the IF along a ray from R136 into IF2 is $n_e = 280\pm50$ cm$^{-3}$. Applying Eq. \ref{eq:dr} with the same assumption used with IF1, we find $Q_0 = 2.0 \times10^{51}$ s$^{-1}$.

IF4 is a bright rim lying much further out in the nebula, at a projected distance 53 pc (220 arcsec) to the east of R136 on the far side of the large faint low-density region that is easily seen in figures \ref{fig:MapHa}-\ref{fig:MapS2}. This large wall is remarkably homogeneous in density and has a thickness between 3.5 and 4 arcsec all along its projected length of 33 pc (140 arcsec). The limb shows up as a density enhancement, with $n_e$ = 125$\pm$30 cm$^{-3}$. It has a nearly constant projected distance from R136. Using the previous technique we find $Q_0 = 4.6 \times 10^{51}$ s$^{-1}$. 

The above estimates of $Q_0$ are all in reasonable agreement with each other and also with  $Q_0 =  4.2\times10^{51}$ s$^{-1}$ estimated by \citet{Crowther_Dessart1998} from adding up the contributions expected from the individual stars in the cluster. These results support the conclusion that the central cluster of stars dominates the ionizing radiation field throughout most of the volume of 30 Dor, and that isolated stars provide a  negligible contribution out to 100pc.

However, there is at least one region, to the SW of R136, where the observed structure indicates ionization by a different source. The IFs designated 1* through 5* on Figure \ref{fig:IFsFinder} show PAH emission closer to R136 than the [S~II] emission. The expected stratification of PAH and H$^+$ emission are consistent with photo-ionization by stars in the region of the OB association LH 99 region. These IFs form an elongated shell structure with a diameter of 15 pc around that cluster. If this interpretation is correct, the emission-line spectrum produced from a different ionizing SED would be different from the bulk of the emission from 30 Dor. Unfortunately, our spectroscopic survey does not cover this particular region, so we cannot make this test.

Interestingly there are examples of regions that appear in the optical data sets as IFs which have no associated PAH emission. One such example is a shell-like structure located at RA = 05$^h$39$^m$4.0$^s$, Dec = -69$^{\circ}$06'07.0'' (J2000) (x=326.0, y=60.0). It has been labeled in figure 14 as the No PAH Shell. An offset dashed line indicates the location. This large feature is difficult to detect in H$\alpha$ images, but is present in the [S II]/H$\alpha$ or [S\ II]/[O\ III] ratio images as a prominent bubble-like structure north of IF 13. However there is no associated feature seen in the 8$\micron$ Spitzer images at this location. This cannot be due to obscuration, which would affect the optical data more severely. Unfortunately this region is just beyond the boundary of our spectrophotometric survey and we can draw no further conclusions about it.

\subsection{Overall symmetry of the ionization structure}

The various surface brightness and ionization maps shown in Figures \ref{fig:MapHa} to \ref{fig:MapS2} all have roughly circular symmetry. The center in each case is at a point offset about 1 arcmin to the E of R136. While at first sight this might seem counter to the idea that R136 marks the center of the source of ionizing radiation, the offset is most likely an artifact of the asymmetrical distribution of gas around R136. The extensive bubble of hot x-ray emitting gas extending off in this direction appears to be a `blow-out' like the ones observed on a smaller scale in M17 (Townsley et al. 2003; Pellegrini et al. 2007) and elsewhere. We conclude that the asymmetrical density distribution has just modestly shifted the apparent center of the ionization structure that otherwise would be centered on R136.

\subsection{30 Dor at a cosmological distance}

In Paper II we will compare the chemical abundances obtained from a point-by-point analysis using our new data set, to the properties that would be derived if 30 Dor were viewed at a cosmological distance and its integrated light analyzed using the `strong-line' analysis techniques (Pagel 1979; Kewley \& Doptia 2002; Denicol{\'o} et al. 2002) generally used in such circumstances.  
  
\begin{figure*}
\center{\includegraphics[scale=0.5]{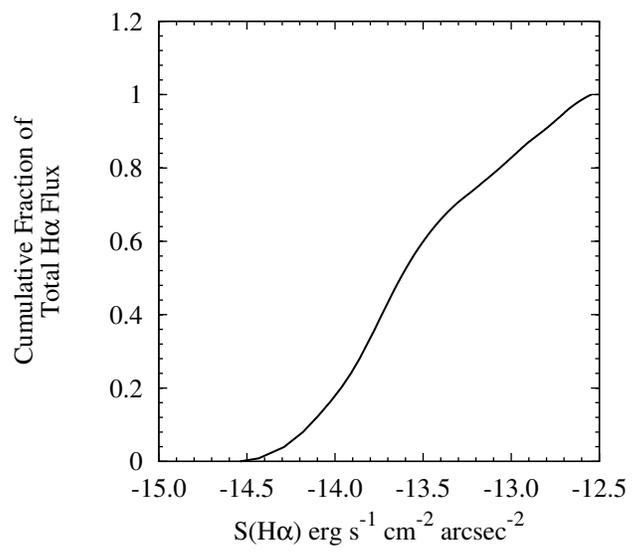}}
\caption{A cumulative histogram of the fraction of total \Ha~ flux from pixels with $\le S(H\alpha)$ from our sky- and continuum-subtracted \Ha~ image over a 12$\times$13 arcmin field of view centered on R136.}
\label{fig:HaHist}
\end{figure*}

It is important to not be misled by the measured properties of the prominent bright arcs that have given 30 Dor the name `Tarantula Nebula'. The majority of the line emission originates in the extended regions of low surface brightness. This is demonstrated in Figure \ref{fig:HaHist}, a histogram of the fraction of nebular flux from regions with a surface brightness $\le S(H\alpha)$ for our 12$\times$13 arcmin$^2$ sky- and continuum-subtracted SOAR \Ha~ image. The bright arcs have a characteristic \Ha~ surface brightness $S ($H$\alpha) \ge 2\times10^{-13}$ erg cm$^{-2}$ s$^{-1}$ arcsec$^{-2}$, while 50 per cent of the integrated \Ha~ emission over our field originates from regions with a surface brightness an order of magnitude fainter than the bright arcs.

\section{Conclusions}
We have obtained and are making publicly available a dense grid of long-slit optical spectra which measure key emission lines in the $\lambda\lambda$4100-7400 \AA\ wavelength region over a 140$\times$80 pc$^2$ (10$\times$6 arcmin$^2$) region of 30 Dor. These spectra were taken at 37 slit positions and then extracted and measured as 4238 individual 1D spectra. The resulting `data cube' of emission-line intensity measurements is also publicly available. Supplementary spectra at a few additional slit positions extended the coverage to a wider wavelength range.

We also obtained a set of subarcsec-resolution direct images in multiple narrow-band filters, covering a 170$\times$180 pc$^2$ (12$\times$13 arcmin$^2$) field of view, which we have used to identify and catalog a large number of structures of special interest. These include regions likely to be locally ionized by embedded stars, as well as edge-on ionization fronts seen in both the visible and infrared and `elephant trunk' pillars. With JWST and ALMA on the horizon and the newly upgraded HST, these well-resolved structures in this nearby object will provide important calibrations for constraining the physical conditions in the PDRs of much more distant, unresolved low-metallicity extragalactic H II regions.

We find that the cluster of O stars centered on R136 is the dominant source of ionization for almost the entire nebula. Our measurements of the [O~III] temperature does not reveal any shock-heated low-ionization gas. The observed emission-line intensity ratios indicate that photoionization is the predominant ionization mechanism. The general radial symmetry of these intensity ratios around a point on the sky near R136 (Figs 9-12) supports the idea that the central cluster is the major source of this photoionization. 

Further evidence includes the absence of strong competition from other potential sources of ionizing radiation within 15 pc of R136 (Sect. 5.1), the large numbers of elephant trunks pointing back towards R136 from points all over 30 Dor (Sect. 5.2), and the variation in the measured thickness of edge-on ionization fronts as a function of distance from R136 (Sect. 5.3). In a follow-up paper we will use these results as the basis for detailed models of the structure of 30 Doradus.

We also showed that the integrated emission-line spectrum of 30 Dor is dominated by the extensive regions of low surface brightness. The bright arcs near R136 catch the eye when one looks at direct images of 30 Dor, but are not what would actually be measured spectroscopically if 30 Dor were so far away that it could not be spatially resolved.

\section{Acknowledgments} 
EWP and JAB gratefully acknowledge support from NSF grant AST-0305833 and NASA grants HST AR-10932 and NNX10AD05G.
JAB gratefully acknowledges additional support from NASA ADP grant NNX10AD05G.
EWP gratefully acknowledges support from NASA grant 07-ATFP07-0124 and partial support from NSF grant AST-0806147. GJF gratefully acknowledges support from NASA grant 07-ATFP07-0124 and from NSF through 0607028 and 0908877.


\clearpage
\begin{deluxetable}{lcccc}
\tablecaption{Summary of imaging observations used in 30 Doradus optical mosaics.}
\tablehead{\colhead{Filter ($\AA$)}&\colhead{No. of Exp. $\times$ Duration}&\colhead{No. Pos.}&\colhead{FWHM (arcsec)}&\colhead{Min/Max Exp Time (s)}\\
\colhead{}&\colhead{at each position}&\colhead{}&\colhead{}&\colhead{}}
\startdata
6563&3 $\times$ 150s&6&0.9&300/1480\\
6738&3 $\times$ 450s&6&1.0&900/5700\\
6850&3 $\times$ 450s&6&0.8-1.0&900/5700\\
5019&3 $\times$ 225s&5&1.0&450/2500\\
5130&3 $\times$ 225s&5&1.0-1.1&450/2500\\
\enddata
\label{tab:SOIsummary}
\end{deluxetable}

\clearpage
\begin{deluxetable}{lcccc}
\tabletypesize{\footnotesize}
\tablecaption{Summary of Blanco and SOAR spectroscopic observations of 30 Doradus}
\tablehead{\colhead{Pos}& \colhead{R.A. (2000)(slit center)}&\colhead{Dec. (2000)}&\colhead{PA(deg E of N)}&\colhead{Exp. Time (s)}}
\startdata
Blanco\\
\hline
1&05:39:16.53&-69:06:37.04&13&405\\
2&05:39:11.25&-69:06:31.18&13&425\\
3&05:39:06.29&-69:06:29.68&13&405\\
4&05:39:00.44&-69:06:22.80&13&250\\
5&05:38:56..29&-69:06:11.60&13&500\\
6&05:38:52.21&-69:06:11.81&13&260\\
7&05:38:45.00&-69:06:03.65&13&260\\
8&05:38:41.38&-69:05:55.66&13&320\\
9&05:38:37.98&-69:05:53.44&13&300\\
10&05:38:35.17&-69:05:48.71&13&120\\
11&05:38:30.08&-69:05:38.24&13&400\\
12&05:38:24.82&-69:05:29.57&13&450\\
13&05:38:17.42&-69:05:28.52&13&500\\
14&05:38:08.61&-69:05:18.33&13&300\\
15&05:37:59.70&-69:04:56.19&13&250\\
16&05:38:49.47&-69:01:28.18&13&600\\
17&05:38:37.48&-69:08:52.47&13&600\\
20&05:39:16.50&-69:04:51.76&103&400\\
21&05:38:28.34&-69:03:50.99&103&400\\
22&05:39:15.65&-69:05:28.77&103&405\\
23&05:38:27.16&-69:04:29.19&103&308\\
24&05:39:12.49&-69:05:54.62&103&500\\
25&05:38:25.24&-69:04:55.67&103&355\\
26&05:39:12.16&-69:06:22.34&103&500\\
27&05:38:25.90&-69:05:23.89&103&460\\
28&05:39:06.04&-69:06:36.65&103&600\\
29&05:38:22.74&-69:05:43.05&103&520\\
30&05:39:07.60&-69:07:05.02&103&700\\
31&05:38:24.16&-69:06:13.86&98&450\\
32&05:39:06.61&-69:07:26.38&103&900\\
33&05:38:21.35&-69:06:31.58&103&500\\
34&05:39:06.01&-69:07:58.72&103&400\\
35&05:38:20.45&-69:07:00.61&103&300\\
36&05:39:01.11&-69:08:34.64&103&600\\
37&05:38:15.70&-69:07:38.32&103&450\\
\hline
SOAR\\
\hline
40&05:38:46.430&-69:05:35.06&10&1000\\
41&05:39:03.142&-69:06:40.45&33.9&3000\\
42&05:39:02.763&-69:06:45.82&61&2000\\
43&05:39:21.240&-69:06:55.44&61&300\\
44&05:39:08.649&-69:07:02.80&80&3000\\
\enddata
\label{tab:SPECpointings}
\tablecomments{Columns 1-5 are position number corresponding to Figures \ref{fig:HaMosaic} and \ref{fig:S2Ha_ratio}, RA and Declination of the slit center, PA and total integrated exposure time. Positions 1-37 are Blanco observations. 40-44 are SOAR observations.}
\end{deluxetable}

\clearpage
\begin{deluxetable}{llccl}
\tabletypesize{\footnotesize}
\tablecaption{Line ID's and wavelengths}
\tablehead{\colhead{$\lambda_{Obs}$}&\colhead{$\lambda{0}$}&\colhead{ion}&\colhead{f$_{\lambda}$ ($R$ = 3.1)}&\colhead{Data Set}}
\startdata
4106&4101&H I&1.43&SOAR\\
4344&4340&H I&1.35&Blanco+SOAR\\
4366&4363&[O\ III]&1.34&Blanco+SOAR\\
4476&4471&He I&1.30&Blanco+SOAR\\
4800&4800&Continuum&1.19&Blanco+SOAR\\
4866&4861*&H I&1.16&Blanco+SOAR\\
4926&4922&He I&1.14&SOAR\\
4963&4959&[O\ III]&1.13&Blanco+SOAR\\
5012&5007*&[O\ III]&1.12&Blanco+SOAR\\
5625&5625&Continuum&0.97&Blanco+SOAR\\
5880&5875*&He I&0.93&Blanco+SOAR\\
6308&6300&[O I]&0.86&SOAR\\
6318&6312&[S\ III]&0.86&Blanco+SOAR\\
6364&6364&NS [O I]&-&Blanco\\
6552&6548&[N\ II]&0.82&Blanco+SOAR\\
6570&6563&H I&0.82&Blanco+SOAR\\
6590&6584*&[N\ II]&0.81&Blanco+SOAR\\
6684&6678&He I&0.80&Blanco+SOAR\\
6721&6716*&[S\ II]&0.79&Blanco+SOAR\\
6738&6731*&[S\ II]&0.79&Blanco+SOAR\\
7072&7065*&He I&0.74&Blanco+SOAR\\
7142&7135*&[Ar III]&0.73&Blanco+SOAR\\
7288&7281&He I&0.71&SOAR\\
7325&7320&[O II]&0.70&SOAR\\
7337&7330&[O II]&0.70&SOAR\\
7758&7751&[Ar III]&0.63&SOAR\\
9075&9069&[S\ III]&0.48&SOAR\\
\enddata

\label{tab:LineID}
\tablecomments{Measured strengths for each of these lines are listed in Table \ref{tab:BlancoFluxes} and/or \ref{tab:SOARFluxes} at every extracted point in the nebula. Rest frame wavelengths with an asterisk indicate the lines used to fit the Blanco data to models at each point in the nebula, as described in Paper II.}

\end{deluxetable}

\clearpage
\begin{deluxetable}{lcccccccccccc}
\tabletypesize{\footnotesize}
\setlength{\tabcolsep}{0.06in}
\tablecaption{Sample of Blanco spectroscopy data cube}
\tablehead{\colhead{1}&\colhead{2}&\colhead{3}&\colhead{4}&\colhead{5}&\colhead{6}&\colhead{7}&\colhead{8}&\colhead{9}&\colhead{10}&\colhead{11}&\colhead{12}&\colhead{13}}
\startdata
Pos&$\Delta$RA&$\Delta$Dec&row&log $n_e$&-err&+err& T(O3)& -err&+err&A$_V$& F(\Hb)& I(\Hb)\\
\hline
1&2&3&4&5&6&7&8&9&10&11&12&13\\
\hline
1&148.1&-181.6&3&1.41&0.41&0.27&13600&1010&1500&1.12&1.94E-15&6.56E-15\\
1&148.6&-179.1&8&1.70&0.24&0.16&10900&1280&3210&1.13&1.88E-15&6.38E-15\\
\nodata&\nodata&\nodata&\nodata&\nodata&\nodata&\nodata&\nodata&\nodata&\nodata&\nodata&\nodata&\nodata\\
37&3.9&-128.5&603&1.88&0.11&0.09&13900&1320&2280&2.86&2.70E-15&6.05E-14\\
38&0.0&0.0&1&2.11&0.001&0.001&10800&8&8&0.00&6.22E-14&6.21E-14\\
38&0.0&0.0&2&2.08&0.001&0.001&0&4&4&1.28&3.55E-15&1.43E-14\\
38&0.0&0.0&3&2.08&0.001&0.001&10700&5&5&0.00&5.75E-16&5.75E-14\\

\hline
$\sigma_{I(H\beta)}$& 4341& $\sigma_{4341}$& 4363 & $\sigma_{4363}$ & 4471 & $\sigma_{4471}$ & 4800 & 
$\sigma_{4800}$ & 4963 & $\sigma_{4963}$ & 5007 & $\sigma_{5007}$\\
\hline
14 & 15 & 16 & 17 & 18 & 19 & 20 & 21 & 22 & 23 & 24 & 25 & 26 \\
\hline
5.65E-17 & 48.42 & 1.37 & 4.97 & 1.07 & 2.42 & 0.93 & 43.52 & 0.98 & 114.01 & 0.83 & 334.02 & 1.22 \\
7.18E-17 & 44.08 & 1.82 & 2.91 & 1.45 & 0.07 & 1.28 & 48.43 & 1.34 & 117.72 & 1.09 & 357.57 & 1.59 \\
\nodata&\nodata&\nodata&\nodata&\nodata&\nodata&\nodata&\nodata&\nodata&\nodata&\nodata&\nodata&\nodata\\
6.13E-16&70.21&4.31&7.04&2.01&1.37&1.68&16.28&1.47&146.94&0.94&456.58&1.29\\
4.15E-18&49.95&0.01&3.68&0.01&3.96&0.01&44.80&0.01&154.31&0.01&462.57&0.01\\
7.96E-19&40.38&0.01&2.89&0.00&3.49&0.00&34.75&0.00&158.12&0.01&482.18&0.01\\
3.20E-18&50.21&0.01&3.55&0.01&4.09&0.00&35.64&0.01&152.61&0.01&457.36&0.01\\

\tablebreak
5625&$\sigma_{5625}$&5876&$\sigma_{5876}$&6300&$\sigma_{6300}$&6302&$\sigma_{6302}$&6318&$\sigma_{6318}$&6364&$\sigma_{6364}$&6552\\
\hline
27&28&29&30&31&32&33&34&35&36&37&38&39\\
\hline
33.09&0.40&15.48&0.36&28.23&0.39&26.12&0.33&2.26&0.20&3.56&0.28&6.51\\
35.14&0.54&16.54&0.34&30.24&0.36&27.52&0.31&2.98&0.19&3.94&0.26&6.78\\
\nodata&\nodata&\nodata&\nodata&\nodata&\nodata&\nodata&\nodata&\nodata&\nodata&\nodata&\nodata&\nodata\\
14.43&0.38&11.21&0.19&6.19&0.17&4.60&0.13&1.88&0.10&0.13&0.13&4.28\\
25.65&0.00&11.82&0.00&6.05&0.00&4.41&0.00&1.70&0.00&0.53&0.00&4.12\\
27.85&0.00&15.75&0.00&8.97&0.00&6.59&0.00&2.48&0.00&2.31&0.00&6.40\\
22.20&0.00&11.87&0.00&6.21&0.00&4.57&0.00&1.71&0.00&0.57&0.00&4.24\\
\hline
$\sigma_{6318}$&6570&$\sigma_{6570}$&6590&$\sigma_{6590}$&6684&$\sigma_{6684}$&6721&$\sigma_{6721}$&6738&$\sigma_{6738}$&7072&$\sigma_{7072}$\\
\hline
40&41&42&43&44&45&46&47&48&49&50&51&52\\
\hline
0.22&284.89&0.77&16.36&0.28&2.02&0.20&25.73&0.29&18.31&0.27&2.21&0.18\\
0.21&284.89&0.67&16.15&0.25&2.51&0.19&23.84&0.26&17.31&0.24&2.71&0.17\\
\nodata&\nodata&\nodata&\nodata&\nodata&\nodata&\nodata&\nodata&\nodata&\nodata&\nodata&\nodata&\nodata\\
0.10&284.72&0.35&9.94&0.12&2.97&0.1&14.12&0.12&10.42&0.11&1.87&0.08\\
0.00&284.85&0.01&10.62&0.00&3.07&0.00&9.51&0.00&7.36&0.00&2.44&0.00\\
0.00&431.16&0.01&16.58&0.00&4.81&0.00&15.4&0.00&11.87&0.00&4.06&0.00\\
0.00&284.87&0.01&10.91&0.00&3.11&0.00&9.87&0.00&7.59&0.00&2.44&0.00\\

\tablebreak
7142& $\sigma_{7142}$\\
\hline
53& 54\\
\hline
9.4& 0.24\\
\nodata&\nodata\\
7.65&0.10\\
10.37&0.00\\
17.56&0.00\\
10.44&0.00\\
\enddata
\label{tab:BlancoFluxes}

\tablecomments{The complete version of this table, with 4241 rows of data, can be found in the electronic version of the journal. The units of electron density and temperature are cm$^{-3}$ and K, respectively. The dereddened \Hb~ surface brightness are reported in units of erg s$^{-1}$ cm$^{-2}$ arcsec$^{-2}$. The strengths of the other emission lines are given in units of 100$\times$S(line)/S(\Hb). The  entries for position number 38 are average values for the whole data set, as described in the text.}

\end{deluxetable}

\clearpage
\begin{deluxetable}{lccccccccccc}
\tabletypesize{\footnotesize}
\setlength{\tabcolsep}{0.06in} 
\tablehead{\colhead{1}&\colhead{2}&\colhead{3}&\colhead{4}&\colhead{5}&\colhead{6}&\colhead{7}&\colhead{8}&\colhead{9}&\colhead{10}&\colhead{11}&\colhead{12}}
\tablecaption{Sample of SOAR data spectroscopy data set}
\startdata
Pos&$\Delta$RA&$\Delta$Dec&row&log n$_e$&-err&+err&T(O3)&-err&+err&T(S3)&-err\\
\hline
1&2&3&4&5&6&7&8&9&10&11&12\\
\hline
40&4.65&-72.58&260&2.34&0.06&0.05&13100&582&728&11100&603\\
40&5.03&-70.42&275&2.43&0.08&0.08&14100&1120&1720&9250&1000\\
40&5.41&-68.26&290&2.67&0.06&0.06&11500&1050&1890&10900&917\\
\nodata&\nodata&\nodata&\nodata&\nodata&\nodata&\nodata&\nodata&\nodata&\nodata&\nodata&\nodata\\
\hline
+err&A$_V$&F(\Hb)&I(\Hb)&$\sigma_{I(H\beta)}$&4106&$\sigma_{4106}$&4344&$\sigma_{4344}$&4368&$\sigma_{4368}$&4474\\
\hline
13&14&15&16&17&18&19&20&21&22&23&24\\
\hline
775&1.49&3.78E-14&1.10E-13&5.80E-16&24.646&1.631&53.838&1.009&6.897&0.851&2.777\\
1960&1.89&1.87E-14&8.13E-14&7.49E-16&32.489&3.662&53.265&2.086&8.330&1.869&3.477\\
1400&1.88&2.08E-14&9.07E-14&7.60E-16&34.825&3.258&53.366&1.866&4.986&1.650&2.839\\
\nodata&\nodata&\nodata&\nodata&\nodata&\nodata&\nodata&\nodata&\nodata&\nodata&\nodata&\nodata\\
\hline
$\sigma_{4474}$&4800&$\sigma_{4800}$&4926&$\sigma_{4926}$&4963&$\sigma_{4963}$&5012&$\sigma_{5012}$&5522&$\sigma_{5522}$&5542\\
\hline
25&26&27&28&29&30&31&32&33&34&35&36\\
\hline
0.676&24.593&0.622&0.793&0.267&166.054&0.547&502.391&0.830&3.189&0.105&2.360\\
1.456&47.975&1.293&3.382&0.548&169.743&0.902&510.836&1.287&5.577&0.202&3.876\\
1.289&45.377&1.151&0.228&0.486&168.095&0.827&512.208&1.196&5.368&0.181&4.101\\
\nodata&\nodata&\nodata&\nodata&\nodata&\nodata&\nodata&\nodata&\nodata&\nodata&\nodata&\nodata\\
\tablebreak
$\sigma_{5542}$&5625&$\sigma_{5625}$&5881&$\sigma_{5881}$&6306&$\sigma_{6306}$&6318&$\sigma_{6318}$&6552&$\sigma_{6552}$&6570\\
\hline
37&38&39&40&41&42&43&44&45&46&47&48\\
\hline
0.096&10.438&0.196&12.476&0.133&0.948&0.064&1.501&0.067&2.560&0.064&285.000\\
0.185&16.704&0.372&12.888&0.236&0.415&0.116&0.966&0.119&2.360&0.111&285.000\\
0.166&17.714&0.334&13.428&0.215&0.703&0.105&1.457&0.109&2.257&0.100&285.000\\
\nodata&\nodata&\nodata&\nodata&\nodata&\nodata&\nodata&\nodata&\nodata&\nodata&\nodata&\nodata\\
\hline
$\sigma_{6570}$&6590&$\sigma_{6590}$&6684&$\sigma_{6684}$&6722&$\sigma_{6722}$&6736&$\sigma_{6736}$&7072&$\sigma_{7072}$&7144\\
\hline
49&50&51&52&53&54&55&56&57&58&59&60\\
\hline
0.284&7.993&0.081&3.462&0.067&6.128&0.067&5.027&0.065&2.284&0.052&9.834\\
0.394&7.911&0.134&3.414&0.115&5.937&0.109&5.010&0.109&2.436&0.088&9.705\\
0.369&6.979&0.120&3.163&0.104&5.005&0.097&4.697&0.098&2.420&0.080&9.577\\
\nodata&\nodata&\nodata&\nodata&\nodata&\nodata&\nodata&\nodata&\nodata&\nodata&\nodata&\nodata\\
\hline
$\sigma_{7144}$&7288&$\sigma_{7288}$&7325&$\sigma_{7325}$&7337&$\sigma_{7337}$&7758&$\sigma_{7758}$&9074&$\sigma_{9074}$\\
\hline
61&62&63&64&65&66&67&68&69&70&71\\
\hline
0.067&0.572&0.044&1.181&0.043&1.403&0.044&1.973&0.044&18.540&0.102\\
0.105&0.753&0.075&1.264&0.073&1.459&0.071&2.137&0.073&17.843&0.145\\
0.096&0.613&0.068&1.225&0.066&1.304&0.063&2.066&0.067&18.485&0.134\\
\nodata&\nodata&\nodata&\nodata&\nodata&\nodata&\nodata&\nodata&\nodata&\nodata&\nodata&\\
\enddata
\label{tab:SOARFluxes}
\tablecomments{The \Hb~ fluxes are in units of erg s$^{-1}$ cm$^{-2}$ arcsec$^{-2}$. T(S3) is the temperature derived from the [S~III] $\lambda$ 9069 / [S~III] $\lambda$ 6312 ratio in a manner identical to that described for the [O~III] temperature. The full table, with 500 rows of data, is available in electronic format upon request.}
\end{deluxetable}

\clearpage
\begin{deluxetable}{lccccccc}
\tabletypesize{\footnotesize}
\tablecaption{Regions of possible local enhancement of ionization due to nearby massive stars}
\tablehead{\colhead{Object ID}& \colhead{RA}&\colhead{Dec}&\colhead{$\Delta$RA} & \colhead{$\Delta$Dec}&\colhead{radius}&\colhead{Spec. Type}&\colhead{Ref.}\\
\colhead{}& \colhead{}&\colhead{}&\colhead{(arcsec)} & \colhead{(arcsec)}&\colhead{(arcsec)}&\colhead{}&\colhead{}}

\startdata
1&05:39:03.44&-69:06:35.72&113&-32&3.97&B0.2V$^{1}$\\
2&05:39:10.17&-69:06:22.63&149&-19&3.97&\nodata\\
3&05:38:57.10&-69:06:06.62&79&-3&1.43&WN6h+O$^{2}$\\
4&05:39:05.43&-69:04:16.27&124&108&2.85&O9.5III$^{1}$\\
5&05:39:05.51&-69:04:31.61&124&92&2.77&O6.5III$^{1}$\\
6&05:38:24.46&-69:07:57.28&-97&-113&2.78&\nodata\\
7&05:38:38.86&-69:08:16.05&-19&-132&4.76&BN6 Iap$^{3}$\\
8&05:38:01.61&-69:04:50.49&-219&73&1.69&\nodata\\
9&05:38:01.31&-69:04:50.25&-221&74&1.35&\nodata\\
10&05:38:41.19&-69:02:57.75&-7&186&3.07&O7V$^{1}$\\
11&05:38:45.39&-69:02:50.84&16&193&3.83&O4V$^{1}$\\
12&05:38:13.98&-69:07:47.63&-153&-104&10.11&unkown\\
13&05:39:20.55&-69:06:54.42&205&-51&4.06&\nodata\\
14&05:39:03.36&-69:09:33.58&113&-210&3.02&\nodata\\
15&05:38:45.08&-69:08:08.06&14&-124&6.26&B1Ia$^{1}$+unknown\\
16&05:38:52.82&-69:06:12.01&56&-8&3.81&O5.5V$^{1}$\\
17&05:38:55.79&-69:05:24.90&72&39&2.41&unknown\\
18&05:38:04.84&-69:07:34.84&-202&-91&3.54&\nodata\\
19&05:38:31.75&-69:02:14.28&-57&230&3.59&unknown\\
20&05:38:41.76&-69:01:58.79&-4&245&4.27&unknown\\
21&05:38:53.54&-69:02:00.56&60&243&2.43&WN6$^{4}$\\
22&05:39:11.63&-69:02:01.02&157&243&3.07&WR$^{5}$\\
23&05:38:15.64&-69:04:37.64&-144&86&9.94&unknown\\
24&05:38:54.94&-69:08:45.76&67&-162&3.77&B0.5Ia$^{1}$\\
25&05:38:56.83&-69:08:42.47&77&-159&4.25&A2/3Ia$^{1}$+unknown\\
26&05:38:54.70&-69:07:45.81&66&-102&5.29&G8V$^{1}$\\
27&05:38:36.43&-69:06:58.70&-32&-55&2.98&WN$^{5}$\\
28&05:38:36.06&-69:06:47.52&-34&-44&2.04&O8.5V$^{1}$+B0.5III$^{1}$+B1Ia$^{2}$\\
29&05:38:51.32&-69:06:41.99&48&-38&2.84&unknown\\
30&05:38:49.79&-69:06:44.12&40&-40&1.7&B0.5I$^{1}$\\
31&05:38:57.40&-69:07:10.75&81&-67&2.73&B3Ia$^{1}$\\
32&05:38:46.60&-69:04:28.07&22&96&2.3&O4V$^{1}$\\
33&05:38:36.91&-69:05:08.19&-30&56&5.97&\nodata\\
34&05:39:01.05&-69:06:30.16&100&-26&2.44&\nodata\\
35&05:38:58.80&-69:05:24.55&88&39&2.46&\nodata\\
36&05:38:17.62&-69:05:42.83&-133&21&13.67&\nodata\\
37&05:39:22.86&-69:07:46.88&217&-103&3.39&\nodata\\
38&05:39:12.35&-69:06:02.74&161&1&4.78&\nodata\\
39&05:38:14.88&-69:04:31.80&-148&92&2.31&\nodata\\
40&05:38:08.53&-69:05:44.36&-182&19&15.11&\nodata\\
41&05:38:10.64&-69:06:17.52&-171&-14&3.69&\nodata\\
42&05:38:09.46&-69:06:22.26&-177&-18&3.69&\nodata\\
43&05:37:49.22&-69:06:14.27&-286&-10&4.41&\nodata\\
44&05:38:24.70&-69:07:43.55&-95&-100&3.06&\nodata\\
45&05:38:30.51&-69:06:46.21&-64&-42&3.07&\nodata\\
46&05:38:29.73&-69:06:57.41&-68&-54&3.07&\nodata\\
47&05:38:31.02&-69:06:37.47&-61&-34&2.63&\nodata\\
\enddata
\label{tab:LocalSources}
\tablecomments{Stellar spectral types are from $^{1}$Bosch et al 1999, $^{2}$Schnurr et al. 2009, $^{3}$Wolborn \& Blades 1997, $^{4}$Robert et al. 2003, $^{5}$Breysacher 1981. Objects found in surveys with no reported spectral type are listed as unknown. Objects with no known counterpart are left blank.}

\end{deluxetable}

\clearpage

\begin{deluxetable}{lcccccc}
\tabletypesize{\footnotesize}
\tablecaption{A catalog of bright, dense pillars and protruding IFs}
\tablehead{\colhead{Pillar ID}& \colhead{RA (J2000)}&\colhead{Dec}&\colhead{$\Delta$RA} & \colhead{$\Delta$Dec}&Length&PA\\
\colhead{}& \colhead{}&\colhead{}&\colhead{(arcsec)} & \colhead{(arcsec)}&\colhead{(arcsec)}&\colhead{}}

\startdata
1&05:37:28.35&-69:01:51.86&-400&252&37&216\\
2&05:37:30.68&-69:04:12.44&-384&111&17&120\\
3&05:37:32.85&-69:04:16.79&-373&107&19&115\\
4&05:37:42.14&-69:02:30.70&-325&213&26&209\\
5&05:37:42.85&-69:04:42.79&-319&81&10&176\\
6&05:37:42.95&-69:04:34.28&-319&90&10&183\\
7&05:37:50.28&-69:06:02.92&-282&1&19&155\\
8&05:37:54.67&-69:05:09.56&-255&54&9&167\\
9&05:37:58.57&-69:03:44.17&-233&140&7&182\\
10&05:37:59.29&-69:05:49.22&-233&15&8&163\\
11&05:38:03.81&-69:05:22.05&-207&42&10&164\\
12&05:38:05.74&-69:06:36.40&-196&-33&6&138\\
13&05:38:08.17&-69:06:51.16&-185&-47&33&167\\
14&05:38:09.49&-69:07:05.29&-180&-61&19&159\\
15&05:38:11.95&-69:07:08.39&-164&-65&26&142\\
16&05:38:13.19&-69:04:25.76&-158&98&13&236\\
17&05:38:14.22&-69:07:00.28&-153&-56&10&131\\
18&05:38:15.89&-69:05:25.19&-142&39&6&260\\
19&05:38:17.03&-69:05:57.00&-137&7&3&112\\
20&05:38:17.56&-69:05:02.66&-131&61&14&200\\
21&05:38:24.12&-69:04:56.88&-99&67&24&206\\
22&05:38:28.52&-69:08:26.14&-72&-142&16&166\\
23&05:38:28.74&-69:03:37.29&-72&147&6&238\\
24&05:38:30.07&-69:08:38.73&-67&-155&24&138\\
25&05:38:33.14&-69:06:02.88&-51&1&5&153\\
26&05:38:34.01&-69:03:33.01&-45&151&12&268\\
27&05:38:34.26&-69:06:26.96&-45&-23&12&153\\
28&05:38:36.05&-69:07:43.52&-35&-100&12&148\\
29&05:38:38.27&-69:08:13.16&-24&-129&9&81\\
30&05:38:40.51&-69:09:58.77&-8&-235&9&111\\
31&05:38:40.97&-69:07:27.27&-8&-83&12&84\\
32&05:38:43.16&-69:07:04.03&3&-60&7&96\\
33&05:38:43.20&-69:01:19.02&3&285&15&173\\
34&05:38:43.23&-69:10:21.48&3&-258&11&132\\
35&05:38:44.16&-69:06:58.89&8&-55&5&90\\
36&05:38:44.25&-69:11:10.68&8&-307&9&127\\
37&05:38:45.18&-69:05:05.74&14&58&8&304\\
38&05:38:45.41&-69:04:19.41&14&104&8&268\\
39&05:38:45.65&-69:10:25.07&19&-261&18&108\\
40&05:38:45.85&-69:09:51.28&19&-227&15&62\\
41&05:38:46.15&-69:07:04.11&19&-60&8&86\\
42&05:38:46.86&-69:09:33.58&25&-210&7&73\\
43&05:38:47.05&-69:07:56.66&25&-113&13&61\\
44&05:38:47.85&-69:07:15.15&30&-71&14&65\\
45&05:38:50.55&-69:10:17.26&46&-253&16&77\\
46&05:38:53.45&-69:08:00.23&57&-116&8&70\\
47&05:38:54.71&-69:05:36.11&68&28&3&343\\
48&05:38:55.01&-69:05:39.94&68&24&3&309\\
49&05:38:55.29&-69:05:40.63&68&23&3&267\\
50&05:38:56.40&-69:07:21.25&73&-77&13&30\\
51&05:38:56.81&-69:05:30.58&78&33&10&297\\
52&05:38:57.29&-69:01:56.18&78&248&5&32\\
53&05:38:57.51&-69:06:27.65&84&-24&16&16\\
54&05:38:57.81&-69:07:56.53&84&-113&19&68\\
55&05:38:58.07&-69:01:51.67&84&252&7&23\\
56&05:38:58.58&-69:02:16.27&89&228&9&233\\
57&05:38:58.84&-69:07:28.60&89&-85&15&43\\
58&05:38:59.48&-69:08:04.34&89&-121&37&67\\
59&05:39:00.01&-69:11:45.42&95&-342&22&83\\
60&05:39:00.14&-69:06:08.15&95&-4&6&340\\
61&05:39:00.29&-69:08:56.01&95&-172&21&52\\
62&05:39:02.13&-69:08:22.12&105&-138&17&46\\
63&05:39:02.47&-69:06:40.10&105&-36&17&29\\
64&05:39:03.49&-69:07:31.02&111&-87&19&33\\
65&05:39:04.13&-69:08:14.74&116&-131&23&57\\
66&05:39:05.07&-69:05:54.83&121&9&7&279\\
67&05:39:07.00&-69:10:49.34&132&-286&7&39\\
68&05:39:07.51&-69:10:26.01&138&-262&14&12\\
69&05:39:08.05&-69:08:47.38&138&-164&18&53\\
70&05:39:11.46&-69:08:15.27&154&-131&9&49\\
71&05:39:11.76&-69:10:53.42&159&-290&14&55\\
72&05:39:12.36&-69:07:44.75&159&-101&13&26\\
73&05:39:13.25&-69:08:56.22&164&-172&9&91\\
74&05:39:14.28&-69:01:45.16&170&259&23&308\\
75&05:39:14.63&-69:10:54.29&175&-290&24&54\\
76&05:39:14.72&-69:07:39.20&175&-95&13&37\\
77&05:39:16.70&-69:05:28.48&186&35&8&350\\
78&05:39:17.98&-69:10:36.02&191&-272&11&32\\
79&05:39:18.95&-69:05:56.61&197&7&11&355\\
80&05:39:18.97&-69:07:45.44&197&-102&8&63\\
81&05:39:19.05&-69:07:45.59&197&-102&8&55\\
82&05:39:21.91&-69:04:31.65&213&92&15&335\\
83&05:39:22.10&-69:05:27.13&213&37&18&341\\
84&05:39:22.91&-69:06:08.13&218&-4&7&359\\
85&05:39:23.79&-69:07:00.26&224&-56&24&14\\
86&05:39:24.53&-69:04:45.88&229&78&9&339\\
87&05:39:26.37&-69:05:59.02&234&5&15&27\\
88&05:39:26.59&-69:04:42.32&240&81&4&344\\
89&05:39:26.62&-69:07:32.13&240&-88&22&28\\
90&05:39:27.52&-69:06:11.28&245&-7&9&349\\
91&05:39:27.79&-69:04:04.10&245&120&28&320\\
92&05:39:30.83&-69:05:55.85&261&8&20&25\\
93&05:39:30.85&-69:10:11.75&261&-248&19&44\\
94&05:39:31.05&-69:05:29.34&261&34&8&341\\
95&05:39:31.14&-69:05:24.43&261&39&6&330\\
96&05:39:31.23&-69:05:20.09&261&44&16&357\\
97&05:39:34.55&-69:05:25.43&283&38&6&324\\
98&05:39:37.13&-69:07:28.57&293&-85&13&12\\
99&05:39:37.64&-69:07:19.20&299&-75&10&13\\
100&05:39:37.86&-69:07:09.53&299&-66&8&8\\
101&05:39:39.19&-69:11:25.06&304&-321&7&240\\
102&05:39:39.53&-69:08:34.63&310&-151&5&29\\
103&05:39:39.73&-69:08:31.87&310&-148&4&58\\
104&05:39:39.77&-69:11:25.05&310&-321&12&258\\
105&05:39:44.91&-69:06:49.21&336&-45&17&2\\
106&05:39:47.81&-69:08:49.79&353&-166&39&19\\
\enddata
\label{tab:Pillars}
\end{deluxetable}

\clearpage
\begin{deluxetable}{lcccccc}
\tabletypesize{\footnotesize}
\tablecaption{Prominent ionization fronts (IFs) suitable for follow-up multi-wavelength studies.}
\tablehead{\colhead{IF ID}& \colhead{RA (J2000)}&\colhead{Dec}&\colhead{$\Delta$RA} & \colhead{$\Delta$Dec}&Length&PA\\
\colhead{}& \colhead{}&\colhead{}&\colhead{(arcsec)} & \colhead{(arcsec)}&\colhead{(arcsec)}&\colhead{}}

\startdata
1&05:38:44.15&-69:06:59.71&9.3&-55.9&31.0&257.7\\
2&05:38:55.02&-69:05:42.02&67.7&21.8&39.8&317.9\\
3&05:38:36.07&-69:06:27.99&-34.1&-24.2&92.2&349.1\\
4&05:39:22.75&-69:07:17.56&216.8&-73.8&87.3&34.7\\
5&05:39:11.80&-69:08:14.33&157.9&-130.5&49.0&68.9\\
6&05:38:59.15&-69:05:14.79&89.9&49.0&38.2&120.0\\
7&05:38:51.70&-69:05:06.06&49.9&57.7&38.2&70.0\\
8&05:37:55.71&-69:05:43.70&-251.1&20.1&75.2&321.9\\
9&05:37:54.88&-69:05:11.42&-255.6&52.4&59.5&306.1\\
10&05:37:45.34&-69:05:14.08&-306.8&49.7&28.5&0.0\\
11&05:38:06.22&-69:07:43.53&-194.6&-99.7&37.3&344.1\\
12&05:38:11.38&-69:08:06.38&-166.9&-122.6&37.3&283.8\\
13&05:39:37.31&-69:07:17.48&295.1&-73.7&28.0&26.0\\
14&05:39:37.57&-69:08:25.22&296.5&-141.4&96.0&63.0\\
15&05:39:23.05&-69:08:33.57&218.4&-149.8&31.0&310.0\\
16&05:39:19.76&-69:08:21.01&200.7&-137.2&14.4&29.2\\
17&05:38:58.60&-69:09:36.36&87.0&-212.6&63.8&16.9\\
18&05:38:32.82&-69:09:17.55&-51.6&-193.8&63.8&340.7\\
19&05:38:58.56&-69:08:43.34&86.8&-159.5&2.4&324.7\\
20&05:39:03.35&-69:08:04.44&112.5&-120.6&13.5&67.9\\
21&05:39:09.79&-69:04:36.61&147.1&87.2&18.3&142.7\\
22&05:38:52.60&-69:04:40.27&54.7&83.5&20.3&128.7\\
23&05:38:44.74&-69:04:19.25&12.5&104.5&20.3&120.0\\
24&05:38:37.67&-69:04:27.59&-25.5&96.2&13.6&45.0\\
25&05:38:39.17&-69:03:31.66&-17.5&152.1&25.9&90.0\\
26&05:38:28.95&-69:02:54.60&-72.4&189.2&29.8&25.7\\
27&05:38:23.40&-69:02:24.23&-102.2&219.6&62.5&111.4\\
28&05:38:12.58&-69:02:16.68&-160.4&227.1&50.4&66.2\\
29&05:38:10.17&-69:02:44.58&-173.4&199.2&11.4&61.0\\
30&05:38:07.60&-69:03:16.01&-187.2&167.8&15.7&39.1\\
31&05:38:14.24&-69:03:29.95&-151.5&153.8&38.8&53.0\\
32&05:38:38.16&-69:00:34.55&-22.9&329.2&28.8&63.8\\
33&05:38:04.92&-69:04:38.91&-201.6&84.9&9.2&26.3\\
34&05:38:34.69&-69:01:05.06&-41.6&298.7&20.8&32.5\\
\hline
1*&05:37:57.50&-69:07:44.65&-241.5&-100.9&107.0&322.2\\
2*&05:38:09.50&-69:08:58.47&-177.0&-174.7&83.8&298.5\\
3*&05:38:19.48&-69:09:55.73&-123.3&-231.9&54.8&343.7\\
4*&05:38:06.23&-69:10:47.75&-194.5&-284.0&54.8&60.8\\
5*&05:37:58.58&-69:08:32.56&-235.7&-148.8&18.4&317.2\\
\enddata
\label{tab:IFs}

\tablecomments{IDs with an asterisk identify IFs with PAH emission closer to R136 than the [S~II] emission which indicate an ionization source other than the central cluster.}

\end{deluxetable}


\end{document}